%% file: main-Hong.tex
\documentclass[conference]{IEEEtran}
\IEEEoverridecommandlockouts
\usepackage{cite}
\usepackage{amsmath,amssymb,amsfonts}
\usepackage{graphicx}
\usepackage{textcomp}
\usepackage{xcolor}
\def\BibTeX{{\rm B\kern-.05em{\sc i\kern-.025em b}\kern-.08em
    T\kern-.1667em\lower.7ex\hbox{E}\kern-.125emX}}

\usepackage{url}
\makeatletter
\def\url@leostyle{%
	\@ifundefined{selectfont}{\def\UrlFont{\sf}}{\def\UrlFont{\small\ttfamily}}}
\makeatother
\usepackage{calc}
\usepackage{bm,bbm}
\usepackage{amsthm}
\usepackage{subfigure}
\usepackage{algorithm,algorithmic}
\usepackage{comment}
\excludecomment{Journal}  
\includecomment{Conf}    

\usepackage{float}
\newtheorem{thm}{\bf{Theorem}}
\newtheorem{prob}{\bf{Problem}}
\newtheorem{cor}{\bf{Corollary}}

\newtheorem{defn}{\bf{Definition}}

\newcommand*{\medcup}{\mathbin{\scalebox{1.25}{\ensuremath{\cup}}}}%
\usepackage{color, xcolor}  

\begin{document}
\title{Pricing Social Visibility Service in Online Social Networks: 
Modeling and Algorithms}

\author{
\IEEEauthorblockN{
Shiyuan Zheng\IEEEauthorrefmark{1}, \,\,\,\,
Hong Xie\IEEEauthorrefmark{2}, \,\,\,\, 
John C.S. Lui\IEEEauthorrefmark{1}
}
\IEEEauthorblockA{ 
\IEEEauthorrefmark{1}\textit{Department of Computer Science and Engineering, 
The Chinese University of Hong Kong} 
\\
\IEEEauthorrefmark{2}\textit{College of Computer Science,
Chongqing University}  
 \\ 
\IEEEauthorrefmark{1}\{syzheng,cslui\}@cse.cuhk.edu.hk,  
\IEEEauthorrefmark{2}xiehong2018@cqu.edu.cn
}
}

\maketitle

\input{0_abstract}


\input{1_introduction}

\input{2_model}

\input{3_analysis}

\input{4_division}

\input{5_evaluation}

\input{6_related}

\input{7_conclusions}

\bibliographystyle{IEEEtran}
\bibliography{myref}



\vspace{12pt}

\end{document}

%% file: 0_abstract.tex

\begin{abstract}  

Online social networks (OSNs) such as 
YoutTube, Instagram, Twitter, Facebook, etc., 
serve as important platforms for users to share their information or content 
to friends or followers.  
Often times, users want to enhance their social visibility, 
as it can make their contents, i.e., opinions, videos, pictures, etc., 
attract attention from more users, 
which in turn may bring higher commercial benefit to them.  
Motivated by this, we propose a mechanism, 
where the OSN operator provides a \textit{``social visibility boosting service''} 
to incentivize ``transactions'' between requesters 
(users who seek to enhance their social visibility 
via adding new ``neighbors'') and suppliers 
(users who are willing to be added as a new ``neighbor'' 
of any requester when certain ``rewards'' is provided).  
We design a posted pricing scheme for the OSN provider to 
charge the requesters who use such boosting service,
and reward the suppliers who contribute to such boosting service.  
The OSN operator keeps a fraction of the payment 
from requesters and distribute the remaining part to participating suppliers 
\textit{``fairly''} via the Shapley value.  
The objective of the OSN provider is to select the 
price and supplier set to maximize 
the total amount of revenue 
under the budget constraint of requesters.  
We first show that the revenue maximization problem is not simpler 
than an NP-hard problem.  
We then decomposed it into two sub-routines, 
where one focuses on selecting the optimal set of suppliers, 
and the other one focuses on selecting the optimal price.  
We prove the hardness of each sub-routine, 
and eventually design a computationally efficient approximation algorithm to solve 
the revenue maximization problem with provable theoretical guarantee 
on the revenue gap.  
We conduct extensive experiments on four public datasets to validate 
the superior performance of our proposed algorithms.  
\end{abstract}

%% file: 1_introduction.tex

\section{\bf Introduction}

OSNs such as YouTube, Instagram, Twitter, Facebook, etc., 
serve as important platforms for users to share their information or content 
to friends or followers, e.g., users in Facebook can share their opinions or status 
to their friends via the friendship network. 
Users in YouTube can share their videos to their subscribers via
the subscriber network.   
A user's friends or followers can further share the information or content 
of this user to their friends or followers.  
Hence, the information or content of a user can be propagated to his 
direct friends or followers, or even multi-hop friends or followers.  
In other words, a user is ``socially visible'' to his direct friends or followers 
and some multi-hop friends or followers.  

Often times, users want to enhance their social visibility, 
as it can make their contents, i.e., opinions, videos, pictures, etc., 
attract attention from more users, 
which in turn may bring higher commercial benefit to them.  
We call a user who wants to enhance social visibility as 
a \textit{requester}.  
One way to enhance a requester's social visibility is to attract some new 
friends or followers.  
It is well known that the OSN is under the ``rich gets richer'' phenomenon, 
which makes it difficult for requesters, 
especially those with low social visibility, 
to attract new friends or followers.  
But when finical incentive is provided, 
some users will be willing to be added as requesters' new friends or followers.   
We call such users as \textit{suppliers}.  
Hence, requesters can provide financial incentives 
to enhance their social visibility.   
This paper aims to answer the following fundamental question: 
\textit{How to set appropriate financial incentives 
to incentivize the ``transaction'' between requesters and suppliers?}  
 
We propose a mechanism, 
where the OSN operator provides a \textit{``social visibility boosting service'' }
to incentivize the transaction between requesters and suppliers.  
The visibility boosting service is sold  
via a posted normalized pricing scheme $(p,q)$, where $p\in [0,1]$ and $q \in [0,1]$.  
Here, $p$ is the price of per unit social visibility 
improvement that the OSN operator 
charges a participating requester, 
and $q$ is the price of per unit contribution 
in improving social visibility of participating requesters 
that the OSN operator pays to a participating supplier.  
We consider the case that each participating supplier adds 
links to all participating requesters and requesters has 
a budget to add $b \in \mathbb{N}_+$ new friends or followers.   
Requesters and suppliers decide whether to participate or not
by comparing their valuations to the posted prices $(p,q)$.  
We consider a proportional transaction fee scheme, 
i.e., the OSN operator keeps a fraction of the payment 
from requesters, and distribute the remaining part (we call it ``reward'') 
to suppliers.  
The objectives of the OSN operator are: 
(1) select the 
price $(p, q)$ and supplier set so to maximize 
the total amount of transaction fees or revenue;  
(2) divide the reward ``fairly'' to all suppliers.  
 
Each of the above two objectives is technically challenging to achieve.  
One can observe that selecting the 
price $(p, q)$ and supplier set involves a mixed optimization problem, 
i.e., with both continuous decision variables, 
i.e., $p$ and $q$, 
and set decision variables, i.e., supplier set.   
This implies that gradient based optimization methods 
does not work for this problem. 
To illustrate the hardness of this optimization problem,  
consider the case where $p$ and $q$ are given 
and our objective is to select the supplier set.  
The number of suppliers may exceed the budget $b$ 
and the OSN operator needs to select at most $b$ suppliers 
from them.  
Due to network externality effect of social visibility, 
suppliers are not independent in enhancing 
the social visibility of a requester.  
In other words, the total improvement of social visibility 
by two suppliers does not equal to 
the total improvement of social visibility 
made by each individual supplier.  
As one will see in Section \ref{sec:algorithm}, 
this dependency among suppliers makes this
simplified problem NP-hard already.  
Also, this dependency among suppliers also 
makes it challenging to fairly divide the reward to suppliers.  
We address these challenges and our contributions are: 

\begin{itemize}
	\item 
	We formulate a mathematical model to quantify social visibility.  
	To the best of our knowledge, we are the first to 
	propose a posted pricing scheme 
	and formulate a revenue maximization problem 
	for visibility boosting service.  
	
	\item 
	We prove that the revenue maximization problem is not simpler 
	than an NP-hard problem.  
	We decomposed it into two sub-routines, 
	where one focuses on selecting the optimal set of suppliers, 
	and the other focuses on selecting the optimal price.  
	We prove the hardness of each sub-problem and propose 
approximation algorithms for each sub-problem with theoretical guarantees 
on the approximation ratio, i.e., a ratio of $(1-1/e)$ for the first sub-routine 
and a ratio which can be make arbitrarily small via increasing the computational 
complexity of searching for the second sub-routine.  
Finally, we prove that combining these approximation 
algorithms, we obtain an algorithm to solve 
the revenue maximization problem with provable theoretical guarantee 
on the revenue gap.  
 
    \item We show how to divide the reward to suppliers fairly via 
    the Shapley value concept.
	We conduct experiments on real-world social network datasets, 
	and the results validate the effectiveness and efficiency of our algorithms. 
	
\end{itemize}

%% file: 2_model.tex

\section{\bf Model \& Problem Formulation}
\label{sec:model}

\subsection{\bf The Social Visibility Model}

\noindent
{\bf Online Social Network.} 
Consider an OSN which is characterized by an unweighted and directed graph 
$\mathcal{G} \triangleq (\mathcal{U}, \mathcal{E})$, 
where $\mathcal{U} \triangleq \{1, \ldots, U\}$ 
denotes a set of $U \in \mathbb{N}_+$ users  
and $\mathcal{E} \subseteq \mathcal{U} \times \mathcal{U}$ 
denotes a set of edges between users.  
Note that a directed edge from user $u \in \mathcal{U}$ 
to user $v \in \mathcal{U}$ is denoted by $(u,v) \in \mathcal{E}$.  
For example, this social network model captures 
the Twitter social network, 
the Facebook social network (each undirected edge between 
user $u$ and $v$ is represented by two 
directed edges $(u,v)$ and $(v,u)$).
We focus on the case that there is no self-loop edges, 
i.e., $(u,u) \notin \mathcal{E}, \forall u \in \mathcal{U}$.  



\noindent
{\bf Social Visibility.} 
Denote a directed path in graph $\mathcal{G}$
as 
\[
\vec{p} \triangleq (u_0\to u_1\to ... \to u_n), 
\] 
where  $(u_i, u_{i+1}) \in \mathcal{E}, \forall i \in \{0,\dots ,n-1 \}$,  
and $u_i \neq u_j, \forall i,j$.   
Note that $u_i \neq u_j, \forall i,j$ captures that 
there is no self-loop edges or circles in the path.  
Denote a set of all directed edges on path $\vec{p}$ as 
\[
\mathcal{F} (\vec{p}) 
\triangleq 
\{(u_0,u_1),(u_1,u_2),...,(u_{n-1},u_n)\}.
\]
Let $L (\vec{p})$ denote the length (or number of hops) 
of path $\vec{p}$, which can be expressed as 
$
L (\vec{p}) 
= 
|\mathcal{F} (\vec{p})|
$.   
Let $\mathcal{P}(u,v ; \mathcal{G})$ denote the set of 
all directed paths (without circles) from user $u$ to user $v$ in $\mathcal{G}$.   
Let $D(u,v; \mathcal{G})$ denote the distance from user $u$ to user $v$ 
in graph $\mathcal{G}$.  
We define $D(u,v; \mathcal{G})$ as the length of the shortest path from $u$ to $v$, i.e., 
\[
D(u,v; \mathcal{G}) 
\triangleq 
\left\{
\begin{aligned} 
& 
\min_{ \vec{p} \in \mathcal{P}(u,v ; \mathcal{G})} 
L (\vec{p}), 
&& 
\text{
if $ \mathcal{P}(u,v ; \mathcal{G}) \neq \emptyset$
} 
\\
& 
+ \infty, 
&& 
\text{
if $ \mathcal{P}(u,v ; \mathcal{G}) = \emptyset$
} .   
\end{aligned}
\right.
\]
Namely, when there is no directed path from $u$ to $v$, 
the distance from $u$ to $v$ is infinite.  
Based on distance between nodes, 
we define user $u$ being $d$-visible to user $v$ 
if 
\[ 
D(v,u ; \mathcal{G}) \leq d.
\]
The $d$-visible set of user $u$ is 
defined as the set of all users to whom user $u$ is $d$-visible, 
formally 
\[
\mathcal{V} (u, d ; \mathcal{G} ) 
\triangleq 
\left\{ 
v \Big| v \in \mathcal{U}, D(v,u ; \mathcal{G}) \leq d 
\right\}.
\]

Let $\tau \in \mathbb{N}_+$ denote the social visibility threshold of an OSN.
This notion of social visibility threshold captures that 
the information of a user can be propagated to its followers 
and its followers may further propagate their own follower 
and so on so forth.
Namely, user $u$ is not visible to users whose   
distance to user $u$ is larger than $\tau$.   
For example, $\tau=1$ models that each user is only visible to its own followers, 
while $\tau=2$ models that each user is visible to its own one-hop and 
two-hop followers. 
Based on $\tau$, we define the notation of visibility.

\begin{figure}[H] 
	\centering 
	\includegraphics[width=0.40\textwidth]{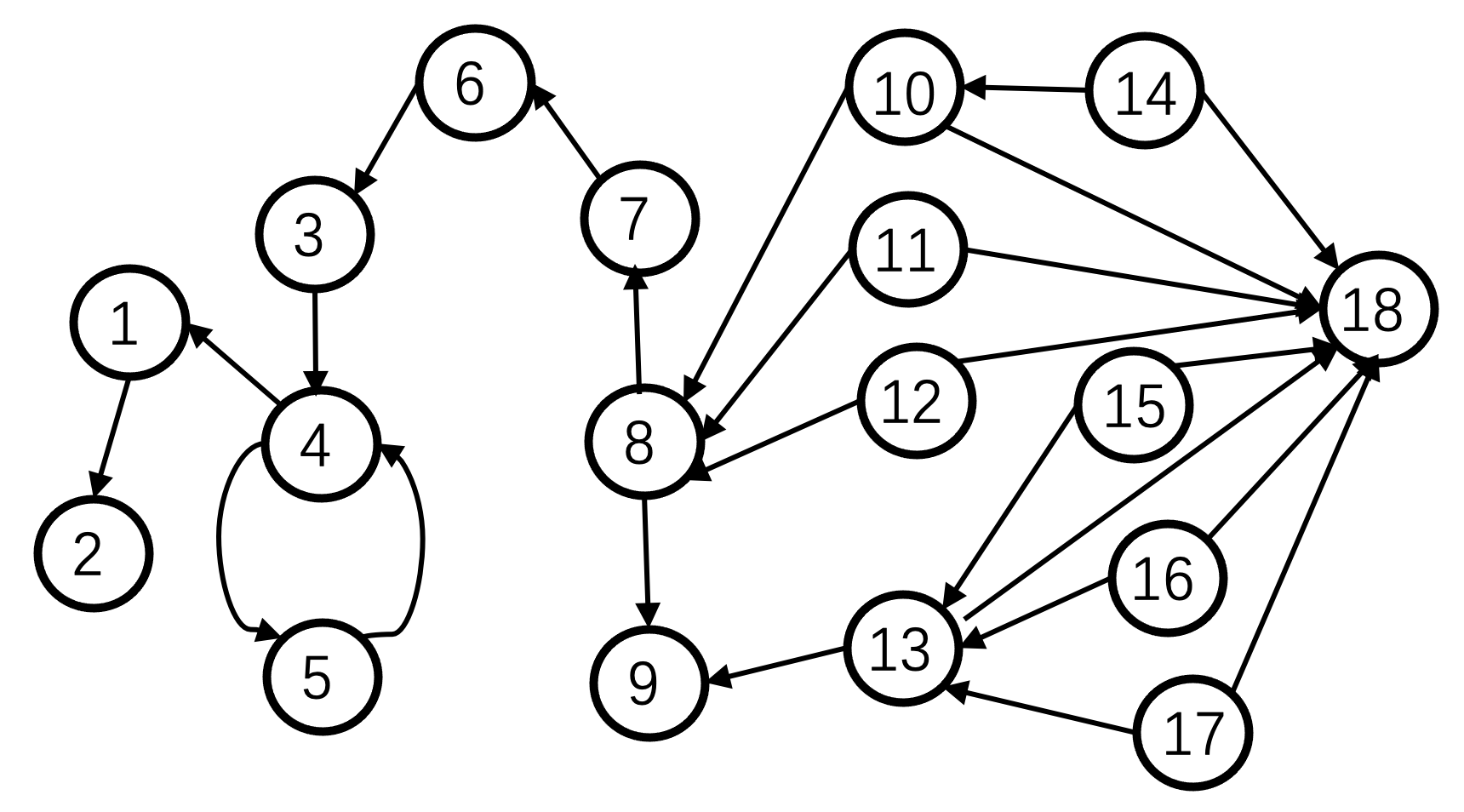} 
	\caption{An example of online social network.} 
	\label{fig:ex1} 
\end{figure}

\begin{defn} 
The visibility of user $u$ 
is the cardinality of his $\tau$-visible set.
	\label{defn:visibility}
\end{defn}
\noindent
For example, Fig.~\ref{fig:ex1} shows that,
user $4$'s $2$-visible set is $\{3,5,6\}$ and  
$3$-visible set is $\{3,5,6,7\}$.
With a social visibility threshold of $\tau=2$, 
user $4$'s visibility is $3$.

\noindent

\subsection{\bf Pricing The Social Visibility}  

\noindent
{\bf The pricing scheme.}  
A \textit{``requester''} is a user in the set $\mathcal{U}$  
who seeks to increase his visibility by 
requesting other users to be his new incoming neighbors.  
Let $\mathcal{R} \subseteq \mathcal{U}$ 
denote a set of all requesters.  
A \textit{``supplier''} is a user in the set $\mathcal{U}$ 
who is willing to be a new incoming neighbor of any requester.  
Let $\mathcal{S} \subseteq \mathcal{U}$ 
denote a set of all suppliers.  
For the ease of presentation, 
we assume that $\mathcal{R} \cap \mathcal{S}  = \emptyset$, 
this captures that a user can not be both requester and supplier.  
We consider the general case that there are some 
users who are neither requesters nor suppliers, 
i.e., $\mathcal{R} \cup \mathcal{S} \subseteq \mathcal{U}$.  
The OSN operator provides a \textit{``social visibility boosting service'' }
to incentivize the ``transaction'' between requesters and suppliers.  
The visibility improvement service is sold at 
via a posted pricing scheme $(p,q)$, where $p\in [0,1]$ and $q \in [0,1]$.  
Here, $p$ is the price of per unit social visibility 
improvement that the OSN operator 
charges a participating requester 
(i.e., requesters who use this social visibility boosting service),
and $q$ is the price of per unit contribution 
in improving social visibility of participating requesters 
that the OSN operator pays to a participating supplier 
(i.e., add links to requesters).  
We consider the case that each participating supplier will add 
links to all participating requesters.   

\noindent
{\bf Requesters' decision model.} 
Each requester $u \in \mathcal{R}$ has a per unit valuation 
(i.e., the per unit price that requester $u$ is willing to pay) 
of $p_u \in [0,1]$ on the improvement of his social visibility.  
A requester will use the social visibility boosting service if 
his per unit valuation is not below the per unit price, 
i.e., $p_u \geq p$.  
Let $\widetilde{R}(p)$ denote a set of all participating requesters 
under price $p$, formally 
\[
\widetilde{R}(p)
\triangleq 
\{
u | p_u \geq p, u \in \mathcal{R}
\}.  
\]
Due to financial budget on boosting service, 
each requester can afford to add at most $b \in \mathbb{N}_+$ 
incoming neighbors.  

\noindent
{\bf Supplier's decision model.}  
Each supplier $u \in \mathcal{S}$ has a per unit valuation 
(i.e., the per unit price that supplier $u$ is willing to participate) 
of $q_u \in [0,1]$ on the per unit contribution 
in improving social visibility of participating requesters.  
Let $\mathcal{M}$ denote a set participating suppliers, 
where $|\mathcal{M}| \leq b$.  
Let $\widetilde{G} (p, \mathcal{M}) $ denote the new graph after 
adding edges from participating suppliers to participating requesters.  
$\widetilde{G} (p, \mathcal{M})$ can be expressed as:  
\[
\widetilde{G} (p, \mathcal{M}) 
= 
(\mathcal{U},  \mathcal{E} \cup (\widetilde{R}(p) \times \mathcal{M}) ).  
\]
Let $I_u (p, \mathcal{M})$ denote the improvement of social visibility 
of requester $u \in \widetilde{R}(p)$.  
It can be expressed as: 
\[
I_u (p, \mathcal{M})
\triangleq 
\left| 
\mathcal{V} (u, \tau ; \widetilde{G} (p, \mathcal{M})  )  
\setminus \mathcal{V} (u, \tau ; \mathcal{G})
\right|.  
\]
Then, the total improvement of social visibility of all participating requesters is 
\[
I (p, \mathcal{M}) 
\triangleq 
\sum_{ u \in \widetilde{R}(p) } I_u (p, \mathcal{M}).  
\]
In other words, all participating suppliers 
make a total contribution of $I (p, \mathcal{M})$ in social visibility enhancement.  
Quantifying $I (p, \mathcal{M})$ among participating suppliers is a non-trivial problem.  
There are two underlying challenges: 
(1) some participating suppliers may have a larger 
number of follower while others may have a small number of followers; 
(2) the network structure poses an externality effect, 
causing the contribution of participating suppliers being correlated.   
To incentivize suppliers to participate, 
one needs to divide $I (p, \mathcal{M})$ fairly among suppliers.  
Let $\phi$ denote a \textit{``fair''} division mechanism, 
which prescribes a contribution denoted by $\phi_u (p, \mathcal{M})$ 
for each participating supplier.  
In order to avoid distracting readers, we defer the 
detail explanation of $\phi$ to the next section.  
Given the fair division mechanism $\phi$, 
a supplier is willing to participant in the social visibility boosting service if 
his per unit valuation does not exceed the per unit price 
that the OSN operator pays, i.e., $q_u \leq q$.  
Let $\widetilde{S}(p)$ denote a set of all potential participating suppliers  
under price $q$, formally 
\[
\widetilde{S}(q)
\triangleq 
\{
u | q_u \leq q, u \in \mathcal{S}
\}.  
\]

\noindent
{\bf Optimal pricing to maximize revenue.}  
Under the posted pricing scheme, 
the total payment made by all requesters to the OSN 
operator can be expressed as: 
\[
G(p, \mathcal{M}) 
\triangleq 
p I (p, \mathcal{M}) .  
\]
The total payment made by the OSN operator 
to all participating suppliers can be expressed as: 
\[
C(q, \mathcal{M}) 
\triangleq
q I (p, \mathcal{M}) .  
\]
The revenue of the OSN operator is therefore:  
\[
R(p,q, \mathcal{M}) 
\triangleq 
G(p, \mathcal{M}) 
- 
C(q, \mathcal{M}).   
\] 
We consider a proportional transaction fee scheme, 
i.e., the OSN operator keeps a fraction $(1-\alpha)$ of $G(p, \mathcal{M}) $, 
where $\alpha \in (0,1)$.  
This is equivalent to imposing $q = \alpha p$.  
We next formulate a revenue maximization problem for the 
OSN operator to select $p,q$ and $\mathcal{M}$.  

\begin{prob}[\bf Optimal pricing to maximize revenue] 
	\label{prob:cons1} 
Select $p,q$ and $\mathcal{M}$ to maximize the revenue of 
the OSN operator:  
	\begin{align*} 
	\max_{ p, q, \mathcal{M} } &  \quad R(p,q, \mathcal{M})  \\ 
	s.t. & \quad \mathcal{M} \subseteq  \widetilde{S}(q), 
	    \left| \mathcal{M} \right|\leq b, \notag  \nonumber \\
	& \quad q = \alpha p,  p \in [0,1], q \in [0,1]. \nonumber
	\end{align*}
\end{prob}

\noindent 
One can observe that the Problem \ref{prob:cons1} is a mixed optimization problem, 
i.e., with both continuous and set decision variables.

%% file: 3_analysis.tex

\section{\bf Algorithms For Optimal Pricing}  
\label{sec:algorithm}  

We first show that Problem \ref{prob:cons1} is not simpler than a 
NP-hard problem.  
Then we decomposed Problem \ref{prob:cons1} 
into two sub-problems.  
We prove the hardness of each sub-problem and propose 
approximation algorithms for each sub-problem.  
Finally, we prove that combining these approximation 
algorithms, we obtain an algorithm to solve 
Problem \ref{prob:cons1} with provable theoretical guarantee 
on the revenue gap.  
\textit{Due to page limit, all technical proofs to lemmas 
and theorems are presented in our supplementary file \cite{TR}.}

\subsection{\bf Hardness Analysis}
\label{sec:hadness}

\noindent
{\bf Hardness analysis.}  
Recall that Problem \ref{prob:cons1} is a mixed optimization problem, 
i.e., with both continuous decision variables, 
i.e., $p$ and $q$, 
and set decision variables, i.e., $\mathcal{M}$.   
This implies that gradient based optimization methods 
does not work for Problem \ref{prob:cons1}.  
To illustrate the hardness of Problem \ref{prob:cons1}, 
we consider a sub-problem of Problem \ref{prob:cons1} 
with given pricing parameters $(p,q)$, 
which is stated as follows:    

\begin{prob}[\bf Optimal supplier set $\mathcal{M}$] 
	\label{prob:sub-problem1} 
Given $p$ and $q$ such that $q = \alpha p$ 
with $0 \leq \alpha \leq 1$,
select $\mathcal{M}$ to maximize the revenue of 
the OSN operator:  
	\begin{align*} 
	\max_{ \mathcal{M} } &  \quad R(p,q, \mathcal{M})  \\ 
	s.t. & \quad \mathcal{M} \subseteq  \widetilde{S}(q), 
	\left| \mathcal{M} \right|\leq b.   \notag  \nonumber 
	\end{align*}
\end{prob}

\noindent
Namely, Problem \ref{prob:sub-problem1} only selects 
the optimal set of suppliers to maximize the revenue of 
the OSN operator.  
Note that the set of potential participating suppliers 
$\widetilde{S}(q)$ is determined as $q$ is given.  
In the following theorem, we analyze its hardness.

\begin{thm}
Problem \ref{prob:sub-problem1} is NP-hard to solve.  
\label{thm:hard:prob2}
\end{thm}

\noindent
Theorem \ref{thm:hard:prob2} states that it is NP-hard to 
locate the optimal set of supplier for Problem \ref{prob:sub-problem1}.  
In other words, it is computationally expensive to locate the 
exact optimal set of supplier for Problem \ref{prob:sub-problem1}.  
Note that Problem \ref{prob:sub-problem1} is a sub-problem of 
Problem \ref{prob:cons1}.  
Namely, Problem \ref{prob:cons1} is harder than 
Problem \ref{prob:sub-problem1}.  
And therefore, locating the exact optimal solution for 
Problem \ref{prob:cons1} is computationally expensive.  
We resort to design approximation algorithms for Problem \ref{prob:cons1} 
with theoretical guarantee.  

\noindent
{\bf Our approach. }  
To address Problem \ref{prob:cons1}, 
we decompose it into two sub-problems, 
which each sub-problem serves as a subroutine.  
In particular, Problem \ref{prob:sub-problem1} 
is the first sub-problem.  
As Theorem \ref{thm:hard:prob2} shows that 
Problem \ref{prob:sub-problem1} is NP-hard, 
we aim to design an approximation algorithm 
which we denote as \texttt{OptSupplierSet}$(p, q)$ 
(its detail is postpone to Section \ref{sec:optSupplier}).  
Algorithm \texttt{OptSupplierSet}$(p, q)$ takes the prices $(p, q)$ 
as an input, and returns an approximately optimal set of suppliers 
under $(p, q)$.  
We use \texttt{OptSupplierSet}$(p, q)$ as 
an oracle to search for the optimal pricing parameter $(p,q)$.  
Formally, we aim to solve the following sub-problem: 

\begin{prob}[\bf Optimal price $p,q$] 
	\label{prob:sub-problem2} 
Given the algorithm \texttt{OptSupplierSet}$(p, q)$, 
select $(p, q)$ so to maximize the revenue of 
the OSN operator:  
	\begin{align*}
	\max_{ p, q} 
	&  \quad R(p,q, \texttt{OptSupplierSet}(p, q))  \\ 
	s.t. & \quad q = \alpha p,  \quad p \in [0,1], q \in [0,1]. \nonumber
	\end{align*}
\end{prob}

\noindent
Note that \texttt{OptSupplierSet}$(p, q)$ returns an approximately 
optimal set of suppliers for each given $(p, q)$, 
Problem \ref{prob:sub-problem2} returns an approximately optimal price.  
We aim to design an approximation algorithm for 
Problem \ref{prob:sub-problem2}, 
which we denote as \texttt{OptPrice}$(\texttt{OptSupplierSet})$ 
(its detail is postpone to Section \ref{sec:optimalpricepq}). 
One needs to supply \texttt{OptPrice}$(\texttt{OptSupplierSet})$ 
with algorithm \texttt{OptSupplierSet}, 
and \texttt{OptPrice}$(\texttt{OptSupplierSet})$ returns 
an approximately optimal $(p,q)$.  
We next proceed to present the design and analysis of 
\texttt{OptSupplierSet}$(p, q)$ and 
\texttt{OptPrice}$(\texttt{OptSupplierSet})$.

\subsection{\bf Design \& Analysis of \texttt{OptSupplierSet}$(p, q)$}  
\label{sec:optSupplier}

\noindent 
{\bf Submodular analysis}. 
First, note that once $p$ and $q$ are given, 
the set of participating requesters $\widetilde{R}(p)$ and 
the set of potential participating suppliers $\widetilde{S}(p)$ 
are given.  
Our objective is to select $\mathcal{M} \in \widetilde{S}(p)$
with the constraint $\mathcal{M} \leq b$, 
so as to maximize the objective function of Problem \ref{prob:sub-problem1}, 
i.e., revenue $R(p,q, \mathcal{M})$, 
where $p$ and $q$ are given and $q = \alpha p$.   
When $q = \alpha p$, the $R(p,q, \mathcal{M})$ can be derived as: 
\begin{align} 
& 
R(p,q, \mathcal{M})  
= 
(1-\alpha)p \sum_{ u \in \widetilde{R}(p) }   I(u, \mathcal{M}) \notag
\\ 
&  
=
(1-\alpha)p 
\sum_{r\in \widetilde{R}(p) } 
\big| 
\medcup_{l \in \mathcal{M} \times \widetilde{R}(p) } \mathcal{V}(l^s,\tau - 1 - D(l^e, r ; \mathcal{G}); \mathcal{G}) \notag
\\
& 
\hspace{0.18in}	
\setminus \mathcal{V}(r, \tau ;\mathcal{G}) \big|, \notag
\end{align}
where $l^s$ and $l^e$ are the start node and the end node 
of  each link  $l\in  \mathcal{M} \times \widetilde{R}(p)$. 
Based on the above closed-form expression of $R(p,q, \mathcal{M})$, 
the following theorem shows the sub-modularity and monotonicity of 
$R(p,q, \mathcal{M})$ with respect to $\mathcal{M}$.  
 
\begin{thm} 
Given $p$ and $q$, such that $q = \alpha p$, 
the revenue $R(p,q, \mathcal{M})$ is monotonously increasing 
and submodular with respect to $\mathcal{M}$.  
\label{thm:submodular}
\end{thm}

\noindent
Theorem \ref{thm:submodular} states that 
$ R(p,q, \mathcal{M})$ is monotonously increasing 
and submodular with respect to $\mathcal{M}$ 
for each given $p$ and $q$ with $q = \alpha p$.

\noindent
{\bf The \texttt{OptSupplierSet}$(p, q)$ algorithm.}  
Based on Theorem \ref{thm:submodular}, 
Algorithm~\ref{alg:greedy} specifies a greedy algorithm to 
implement \texttt{OptSupplierSet}$(p, q)$.  
The core idea of Algorithm~\ref{alg:greedy} is that 
we select suppliers one by one.  
Each time we select the supplier that achieves 
the largest marginal improvement in the revenue.  

\begin{algorithm}[H] 
	\begin{algorithmic}[1]
		\STATE init $\mathcal{M}=\emptyset$
		\FOR{$t=1$ to $b$}
		\STATE $u^\ast \leftarrow  \arg \max_{u\in \widetilde{S}(p)} 
		R(p,q,\mathcal{M}\cup \{ u\}) - R(p,q,\mathcal{M})$
		\STATE $\mathcal{M} \leftarrow \mathcal{M}  \cup \{ u^\ast \}$
		
		\ENDFOR
		\RETURN ${\hat{\mathcal{M}}}^\ast \leftarrow \mathcal{M}$
	\end{algorithmic}
	\caption{\texttt{OptSupplierSet}$(p, q)$ }
	\label{alg:greedy}
\end{algorithm}

The following theorem presents the theoretical guarantees for Algorithm~\ref{alg:greedy}.

\begin{thm}
Given $p$ and $q$, such that $q = \alpha p$.  
The output $\hat{\mathcal{M}^\ast}$ of Algorithm~\ref{alg:greedy} 
satisfies: 
	We have  
	\begin{equation*}
		 R(p,q, \hat{\mathcal{M}}^\ast)  
		\geq 
		\left(
		1 - \frac{1}{e} 
		\right) 
		R(p,q, \mathcal{M}^\ast), 
	\end{equation*}
where $\mathcal{M}^\ast$ denotes the optimal set of suppliers via exhaustive search. 
	\label{thm:greedy_ratio}
\end{thm}

Theorem \ref{thm:greedy_ratio} states that Algorithm~\ref{alg:greedy} 
searches a set of suppliers with approximation ratio of
at least $1 - 1/e$.  

\subsection{\bf Design \& Analysis of \texttt{OptPrice}$(\texttt{OptSupplierSet})$}
\label{sec:optimalpricepq}

\noindent
{\bf Perfect search. }  
Now we consider the case that 
given \texttt{OptSupplierSet}$(p, q)$ outlined in 
Algorithm \ref{alg:greedy}, 
we can obtain exact optimal price $(p, q)$ for 
Problem \ref{prob:sub-problem2}.  
The objective is to understand the impact of 
the approximate optimal supplier set returned by 
\texttt{OptSupplierSet}$(p, q)$ on the optimal price $(p, q)$.  
We state this impact in the following theorem.  

\begin{thm}
Let $(p^\ast_3, q_3^\ast)$ denote the optimal price of 
Problem \ref{prob:sub-problem2}. 
Given \texttt{OptSupplierSet}$(p, q)$ outlined in Algorithm \ref{alg:greedy}, 
we have: 
\[
R(p^\ast_3, q_3^\ast, \hat{\mathcal{M}}^\ast (p^\ast_3, q_3^\ast) ) 
\geq 
\left(
		1 - \frac{1}{e} 
		\right) 
R(p^\ast, q^\ast, \mathcal{M}^\ast),   
\]
where $(p^\ast, q^\ast, \mathcal{M}^\ast)$ 
denotes one optimal solution for the Problem \ref{prob:cons1}.  
\label{thm:algo:optProb3}
\end{thm}

Theorem \ref{thm:algo:optProb3} states that with 
\texttt{OptSupplierSet}$(p, q)$ outlined in Algorithm \ref{alg:greedy}, 
an optimal solution of Problem \ref{prob:sub-problem2} 
attains an approximation ratio of $(1- 1/e)$ of 
the optimal solution of Problem \ref{prob:cons1}.  
However, the optimal solution of Problem \ref{prob:sub-problem2} is 
not easy to obtain.  
One challenge is that the closed-form expression of the objective 
function of Problem \ref{prob:sub-problem2}, 
i.e., $R(p,q, \texttt{OptSupplierSet}(p, q))$, 
is not available, 
and its gradient is not available as well.  
Thus, we resort to gradient-free algorithms to solve 
Problem \ref{prob:sub-problem2}, 
in particular, the discretized search method.  
We leave it as a future work to study other 
advanced methods such as 
Monte Carlo optimization to address this challenge.

\noindent
{\bf Discretized search.}  
Note that $q = \alpha p$.  
We therefore discretize the domain of $p$, i.e., $[0,1]$, uniformly:  
\[
\mathcal{A}(\epsilon) 
\triangleq 
\left\{
0, \epsilon, 2 \epsilon, \ldots, 
\left\lfloor \frac{1}{\epsilon} \right\rfloor \epsilon, 1 
\right\},
\]
where $\epsilon \in (0,1]$  is price search step the                                                                                                                                                                                                                                                                                                                                                                                                                                                                                                                                                                                                                                          
The OSN operator can control $\epsilon$ to adjust the 
number of points in $\mathcal{A}(\epsilon)$.  
The OSN operator can use exhaustive search method to 
select the optimal price $\mathcal{A}(\epsilon) $, 
denoted by $p_{\text{DS}}^\ast$.  
Algorithm \ref{alg:DS} outlines this discretized search algorithm.  
We denote $q_{\text{DS}}^\ast = \alpha p_{\text{DS}}^\ast$.  
The set of supplier is then 
$
\hat{\mathcal{M}}^\ast (p_{\text{DS}}^\ast, q_{\text{DS}}^\ast).  
$  
The following theorem states the theoretical guarantee of 
this method.  

\begin{algorithm}[H] 
	\begin{algorithmic}[1]
		\STATE $\text{Rev} \leftarrow 0$
		\FOR{$p \in \mathcal{A} (\epsilon) $ }
		\STATE $q \leftarrow \alpha p$
		\STATE  $\mathcal{M} \leftarrow$\texttt{OptSupplierSet}$(p, q)$ 
		\IF{ $R(p, q, \mathcal{M}) \geq \text{Rev}$   }
		\STATE $p_{\text{DS}}^\ast \leftarrow p, 
		q_{\text{DS}}^\ast \leftarrow q, 
		\hat{\mathcal{M}}^\ast (p_{\text{DS}}^\ast, q_{\text{DS}}^\ast) 
		\leftarrow \mathcal{M}$
		\STATE $\text{Rev} \leftarrow R(p, q, \mathcal{M})$
		\ENDIF
		
		\ENDFOR
		\RETURN $p_{\text{DS}}^\ast, q_{\text{DS}}^\ast, \hat{\mathcal{M}}^\ast (p_{\text{DS}}^\ast, q_{\text{DS}}^\ast)$
	\end{algorithmic}
	\caption{ \texttt{OptPrice}$(\texttt{OptSupplierSet})$ }
	\label{alg:DS}
\end{algorithm}

\begin{thm}
Suppose $R(p, \alpha p, \hat{\mathcal{M}}^\ast (p, \alpha p) )$  
is $\beta$-Lipschitz with respect to $p$,  
then the output $(p_{\text{DS}}^\ast, q_{\text{DS}}^\ast, 
\hat{\mathcal{M}}^\ast (p_{\text{DS}}^\ast, q_{\text{DS}}^\ast) )$ 
of Algorithm \ref{alg:DS} satisfies   
\[
R(p_{\text{DS}}^\ast, q_{\text{DS}}^\ast, 
\hat{\mathcal{M}}^\ast (p_{\text{DS}}^\ast, q_{\text{DS}}^\ast) )
\geq 
R(p^\ast_3, q_3^\ast, \hat{\mathcal{M}}^\ast (p^\ast_3, q_3^\ast) ) 
- \beta \epsilon.     
\]
\label{thm:algo:DS}
\end{thm}
Theorem \ref{thm:algo:DS} states that when 
$R(p, \alpha p, \hat{\mathcal{M}}^\ast (p, \alpha p) )$  
is $\beta$-Lipschitz with respect to $p$, 
the revenue gap between the discretized search method 
and the perfect search method is 
bounded by $\beta \epsilon$.  
An attractive of this algorithm is that
the OSN operator can make this gap arbitrarily small 
by selection small enough $\epsilon$.  
Selecting smaller $\epsilon$ leads to a larger computational 
complexity.  
Thus, Theorem \ref{thm:algo:DS} serves as a building block 
for an OSN operator to make a tradeoff between 
computational complexity and revenue.  
Combining them all, we prove the revenue gap between 
the pricing searched by the 
discretized search method and the optimal pricing of Problem \ref{prob:cons1}.  

\begin{cor}
Given \texttt{OptSupplierSet}$(p, q)$ outlined in Algorithm \ref{alg:greedy}.  
Suppose $R(p, \alpha p, \hat{\mathcal{M}}^\ast (p, \alpha p) )$  
is $\beta$-Lipschitz with respect to $p$.   
The output $(p_{\text{DS}}^\ast, q_{\text{DS}}^\ast, 
\hat{\mathcal{M}}^\ast (p_{\text{DS}}^\ast, q_{\text{DS}}^\ast) )$ 
of Algorithm \ref{alg:DS} satisfies that 
\[
R(p_{\text{DS}}^\ast, q_{\text{DS}}^\ast, 
\hat{\mathcal{M}}^\ast (p_{\text{DS}}^\ast, q_{\text{DS}}^\ast) )
\geq 
\left(
		1 - \frac{1}{e} 
		\right) 
R(p^\ast, q^\ast, \mathcal{M}^\ast)
- \beta \epsilon.     
\]
\end{cor}

%% file: 4_division.tex

\section{\bf Algorithms for Fair Division of Contribution}
   

Our results so far can locate the approximate optimal price 
and supplier set, i.e., $(p_{\text{DS}}^\ast, q_{\text{DS}}^\ast, 
\hat{\mathcal{M}}^\ast (p_{\text{DS}}^\ast, q_{\text{DS}}^\ast) )$.  
The remaining issue is how to divide the contribution among 
participating suppliers fairly.  
As we mentioned in Section \ref{sec:model} that 
 a \textit{``fair''} division mechanism 
is important to incentivize the participation of suppliers.  
Given the approximate optimal solution 
$(p_{\text{DS}}^\ast, q_{\text{DS}}^\ast, 
\hat{\mathcal{M}}^\ast (p_{\text{DS}}^\ast, q_{\text{DS}}^\ast) )$, 
the OSN operator needs to divide a total contribution of social 
visibility improvement 
$
I (
p_{\text{DS}}^\ast, 
\hat{\mathcal{M}}^\ast (p_{\text{DS}}^\ast, q_{\text{DS}}^\ast)
) 
$  
among all participating suppliers.  
Note that the naive equal division, i.e., participating suppliers 
equally share 
$
I (
p_{\text{DS}}^\ast, 
\hat{\mathcal{M}}^\ast (p_{\text{DS}}^\ast, q_{\text{DS}}^\ast)
)
$
is \textit{not} a fair division.  
This is because:  
(1) some participating suppliers may have a larger 
number of follower while others may have a small number of followers; 
(2) the network structure poses an externality effect, 
causing the contribution of participating suppliers being correlated.  
To achieve fair division, 
we apply the Shapley value \cite{Shapley1951} to divide 
$
I (
p_{\text{DS}}^\ast, 
\hat{\mathcal{M}}^\ast (p_{\text{DS}}^\ast, q_{\text{DS}}^\ast)
)
$.  
Formally, each participating supplier $u \in \hat{\mathcal{M}}^\ast (p_{\text{DS}}^\ast, q_{\text{DS}}^\ast)$ 
gets the following share of profit: 
\begin{align*}
& 
\phi_u (p_{\text{DS}}^\ast, 
\hat{\mathcal{M}}^\ast (p_{\text{DS}}^\ast, q_{\text{DS}}^\ast)) 
\\
& 
= 
\sum_{\mathcal{M}\subseteq {
\hat{\mathcal{M}}^\ast (p_{\text{DS}}^\ast, q_{\text{DS}}^\ast)
} \setminus\{u\}  }  
\frac{|\mathcal{M}|!
(|\hat{\mathcal{M}}^\ast (p_{\text{DS}}^\ast, q_{\text{DS}}^\ast)| 
-|\mathcal{M}|-1)!}{|\hat{\mathcal{M}}^\ast (p_{\text{DS}}^\ast, q_{\text{DS}}^\ast))|!} 
\\
& 
\hspace{0.18in}
\times 
(  
I(p_{\text{DS}}^\ast, 
\mathcal{M}\cup \{u\})
 - 
I(p_{\text{DS}}^\ast, 
\mathcal{M}) 
).  
\end{align*}  
Readers can refer to \cite{Shapley1951} 
for why Shapley value achieves fair division.  

One challenge is that the computational complexity of evaluating 
$\phi_u (p_{\text{DS}}^\ast, 
\hat{\mathcal{M}}^\ast (p_{\text{DS}}^\ast, q_{\text{DS}}^\ast))$ 
is exponential in the cardinality of 
$
\hat{\mathcal{M}}^\ast (p_{\text{DS}}^\ast, q_{\text{DS}}^\ast)
$.   
To address this computational challenge, 
we propose to use the sampling algorithm \cite{Bachrach2010} 
to approximate $\phi_u (p_{\text{DS}}^\ast, 
\hat{\mathcal{M}}^\ast (p_{\text{DS}}^\ast, q_{\text{DS}}^\ast))$.  
Let $\sigma = (u_1,\cdots,u_{\widetilde{b}})$ denote 
an ordering of the participating suppliers, 
where $\widetilde{b} = \min\{b, |\hat{\mathcal{M}}^\ast (p_{\text{DS}}^\ast, q_{\text{DS}}^\ast)|\}$ and 
$u_i \in \hat{\mathcal{M}}^\ast (p_{\text{DS}}^\ast, q_{\text{DS}}^\ast)$ denotes the participating supplier in the $i$-th order.  
Denote the set of players ranked before player $i$ in the order $\sigma$ as
\begin{equation*}
\mathcal{S}_i^\sigma \triangleq \{ \text{all players ranked before } u_i \text{ in the order } \sigma \}. 
\end{equation*}
Based on \cite{Bachrach2010}, 
the 
$\phi_u (p_{\text{DS}}^\ast, 
\hat{\mathcal{M}}^\ast (p_{\text{DS}}^\ast, q_{\text{DS}}^\ast))$  can be rewritten as 
\begin{align}
& 
\phi_u (p_{\text{DS}}^\ast, 
\hat{\mathcal{M}}^\ast (p_{\text{DS}}^\ast, q_{\text{DS}}^\ast)) 
\nonumber \\
& 
= 
\mathbb{E}_{\sigma \sim \text{Uniform}(\Omega)}
[
I(p_{\text{DS}}^\ast, 
\mathcal{S}_i^\sigma \cup \{u\})
 - 
I(p_{\text{DS}}^\ast, 
\mathcal{S}_i^\sigma) 
], 
\label{eq:shapley2}
\end{align}
where $\Omega$ denotes a set of all participating suppliers, 
and $\text{Uniform}(\Omega)$ denotes a uniform distribution over 
$\Omega$.  
Based on Equation \ref{eq:shapley2}, 
Algorithm \ref{algo:shapley} outlines a sampling algorithm 
to approximate $
\phi_u (p_{\text{DS}}^\ast, 
\hat{\mathcal{M}}^\ast (p_{\text{DS}}^\ast, q_{\text{DS}}^\ast)) 
$.  
 
\begin{algorithm}[H] 
	\begin{algorithmic}[1]
		\STATE $\hat{\phi}_u = 0$
		\FOR{$k=1$ to $K$}
		\STATE generate a ordering $\sigma$ uniformly at random from $\Omega$
		\STATE  
		$\hat{\phi}_{u} \leftarrow 
		[ (k-1)\hat{\phi}_{u} + I(p_{\text{DS}}^\ast, 
\mathcal{S}_i^\sigma \cup \{u\})
 - 
I(p_{\text{DS}}^\ast, 
\mathcal{S}_i^\sigma) ] / k $
		\ENDFOR
		\RETURN $\hat{\phi}_{u}$
	\end{algorithmic}
	\caption{Approximating $\phi_u (p_{\text{DS}}^\ast, 
\hat{\mathcal{M}}^\ast (p_{\text{DS}}^\ast, q_{\text{DS}}^\ast)) $}
	\label{algo:shapley}
\end{algorithm}

The following theorem states theoretical guarantee for  
the approximation accuracy of Algorithm~\ref{algo:shapley}. 

\begin{thm} 
	The output $\hat{\phi}_{\sigma_n}$ of Algorithm~\ref{algo:shapley} satisfies
	\begin{align*}
	& 
	| \hat{\phi}_{u}- \phi_u (p_{\text{DS}}^\ast, 
\hat{\mathcal{M}}^\ast (p_{\text{DS}}^\ast, q_{\text{DS}}^\ast))  | 
\\
&	\leq \frac{\max_{\sigma \in \Omega} 
	[ I(p_{\text{DS}}^\ast, 
\mathcal{S}_i^\sigma \cup \{u\})
 - 
I(p_{\text{DS}}^\ast, 
\mathcal{S}_i^\sigma) ]
	 }{\sqrt{K}}  
	\sqrt{\frac{1}{2} \ln \frac{2}{\delta} },
	\end{align*}
	with a probability of at least $1-\delta$, where $\delta \in (0,1]$.
\label{thm:Sahpely}
\end{thm}

Theorem \ref{thm:Sahpely} states that one can control the approximation 
error of Algorithm~\ref{algo:shapley} arbitrarily small 
by selecting a sufficiently large simulation rounds $K$.

%% file: 5_evaluation.tex

\section{\bf Performance Evaluation}
\label{sec:evaluation}
In this section, we conduct experiments on real-world datasets 
to evaluate the performance of our algorithms, 
and results show the their superior performance.


\subsection{\bf Experimental Settings}

\noindent
\textbf{Datasets.} 
We evaluate our algorithms on four public datasets, 
whose overall statistics is summarized in 
Table~\ref{table:datasets}.  

\noindent
\textbf{$\bullet$ Residence \cite{konect}. }
This dataset contains friendship connections between 217 residents, 
who live at a residence hall in the Australian National University campus. 

\noindent
\textbf{$\bullet$ Blogs \cite{konect}.}  
This dataset contains hyperlinks between blogs 
in the context of 2004 US election.  
Blogs are mapped as nodes  and hyperlinks are mapped 
as directed links. 


\noindent
\textbf{$\bullet$ Facebook \cite{snapnets}.} 
It is a page-page network of verified Facebook sites where nodes correspond to official Facebook pages and edges to mutual likes between pages. 

\noindent
\textbf{$\bullet$ DBLP \cite{nawaz2015intra}. }
This dataset contains a sub-network of the co-author 
network of the DBLP network.  
Scholars who have published papers in major conferences (those considered in DBLP) 
are mapped as nodes.  
Each co-author relationship between two scholars is mapped as 
two directed edges between these two scholars with different directions. 

From Table~\ref{table:datasets}, one may 
argue that the scales of the above four datasets are not large.  
We intentionally make this choice because: 
(1) We have already proved the quality gap; 
(2) To compare with baseline algorithms such as the brute force method, 
the OSN has to be small to make it computationally feasible.


\begin{table}[h!]
	\caption{Statistics of four datasets.}	
	\centering
	\begin{tabular}{ |c|c|c|c|c|c|c| } 	
		\hline
		datasets & \#nodes & \#links & type & $\tau$   \\ 
		\hline
		Residence & 217 & 2,672 & directed & 2 \\ 
		Blogs & 1,224 & 19,025 & directed & 2\\ 
		Facebook & 5,908 & 41,729 & undirected & 2\\
		DBLP & 10,000 & 55,734 & undirected & 2\\ 
		\hline
	\end{tabular}
	\label{table:datasets}
\end{table}

\noindent
\textbf{Parameter setting.}
To reflect the real-world setting that only a small portion of 
users in an OSN is interested in the social visibility boosting service, 
we select $\gamma$ fraction of users uniformly at random from 
the user population $\mathcal{U}$ as requesters $\mathcal{R}$, 
and another $\gamma$ fraction of users uniformly at random from 
the user population $\mathcal{U}$ as suppliers $\mathcal{S}$.   
We set $\gamma$ as  0.05, 0.25, 0.5 and 0.1
for dataset Residence, Blogs, Facebook and DBLP respectively.  
Here we select different $\gamma$ on different datasets, 
in order to reveal the impact of size of requesters and 
suppliers on our algorithm.  
For each requesters $u \in \mathcal{R}$, 
valuation $p_u$ is sampled from Beta distribution $B(3,6)$, 
and $q_u$ is sampled from Beta distribution $B(6,3)$.
Throughout the experiments, we fixed $\alpha = 0.6$ for the OSN operator, 
i.e., the OSN operator  keeps 40\% of the payment from the 
requesters as the transaction fee.

\noindent
\textbf{Metrics \& baselines.}
We use $R(p,q,\mathcal{M})$, the revenue of the OSN operator 
as the evaluation metric.  
To understand the accuracy and running time of 
our \texttt{OptSupplierSet} algorithm, i.e., Algorithm \ref{alg:greedy}, 
we compare it with the following two baselines: 
(1) \texttt{Brute}, which selects the optimal set of participating suppliers 
under each given prices $(p,q)$ via exhaustive search; 
(2) \texttt{TopVis}, which selects suppliers with the top-$b$  
social visibility from the potential supplier set under each given prices $(p,q)$.  
To evaluate our \texttt{OptPrice} algorithm, i.e., Algorithm \ref{alg:DS}, 
we vary the search step $\epsilon$ from 
0.2, 0.1, 0.05, 0.025 to 0.0125.

\subsection{\bf Evaluating \texttt{OptSupplierSet}}
\begin{figure}[htb]
	\centering  
	\subfigure[Residence]{
		\includegraphics[width=0.225\textwidth]{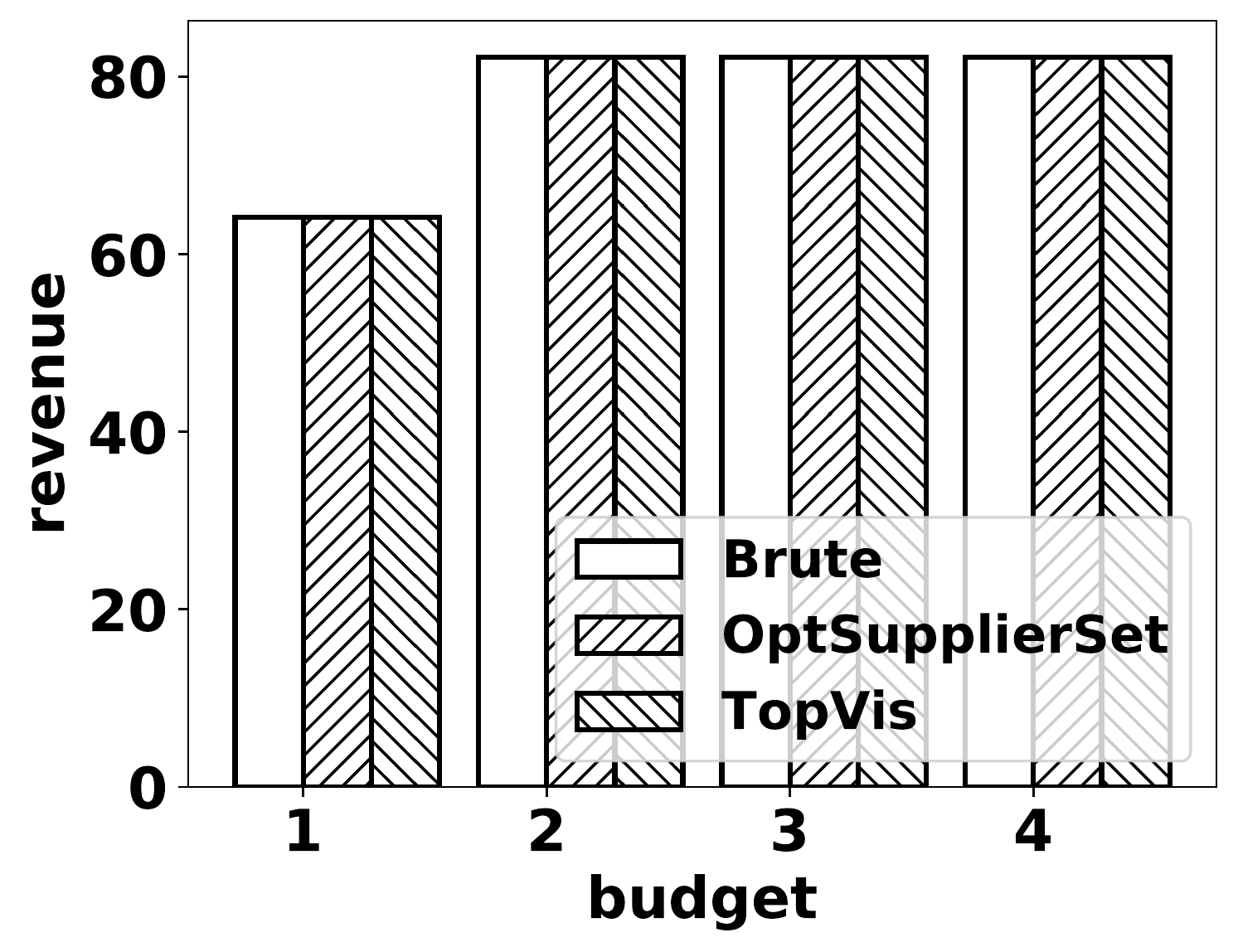}
	}
	\subfigure[Blogs]{
		\includegraphics[width=0.225\textwidth]{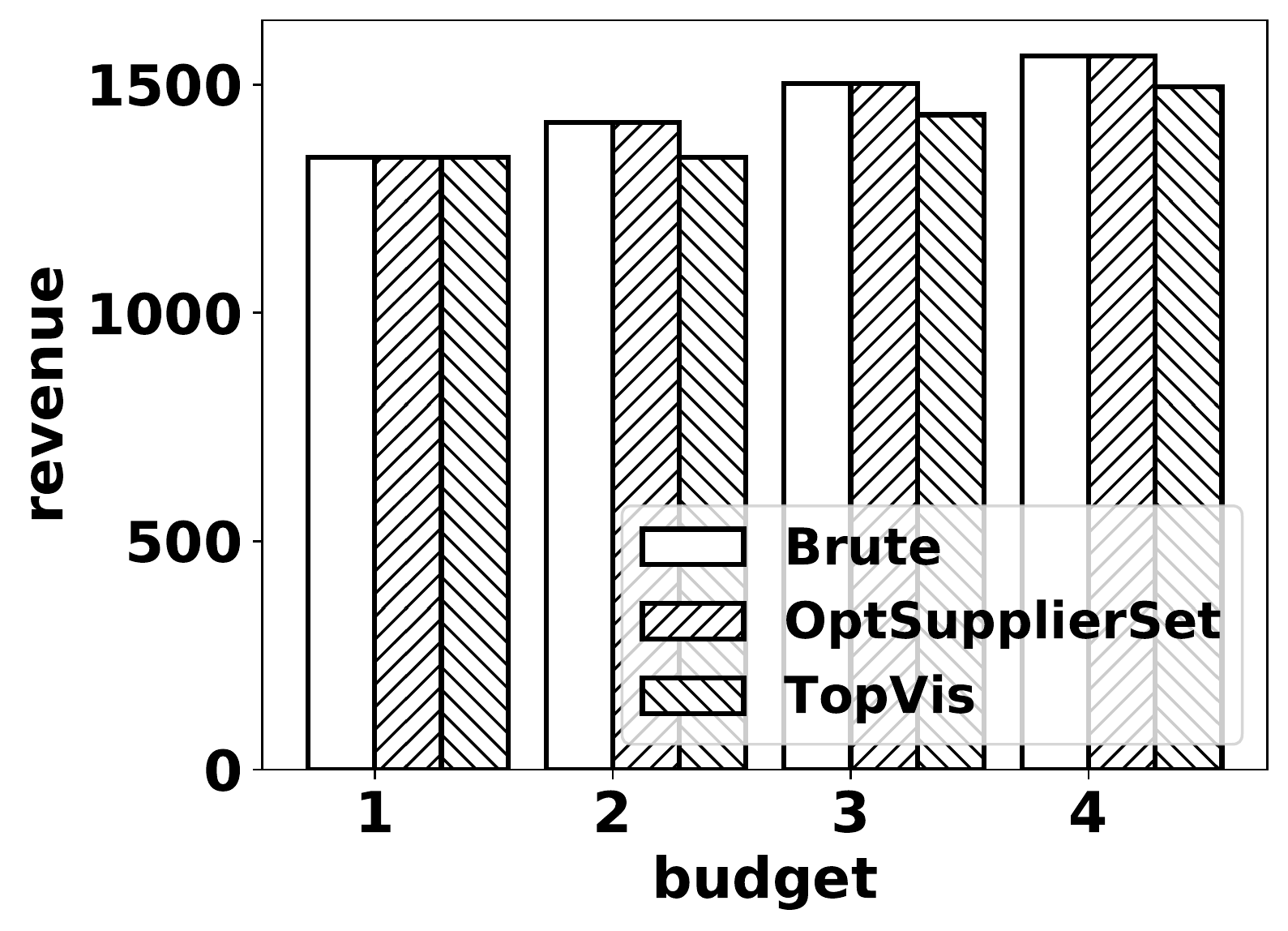}
	}
	\\
	\subfigure[Facebook]{
		\includegraphics[width=0.225\textwidth]{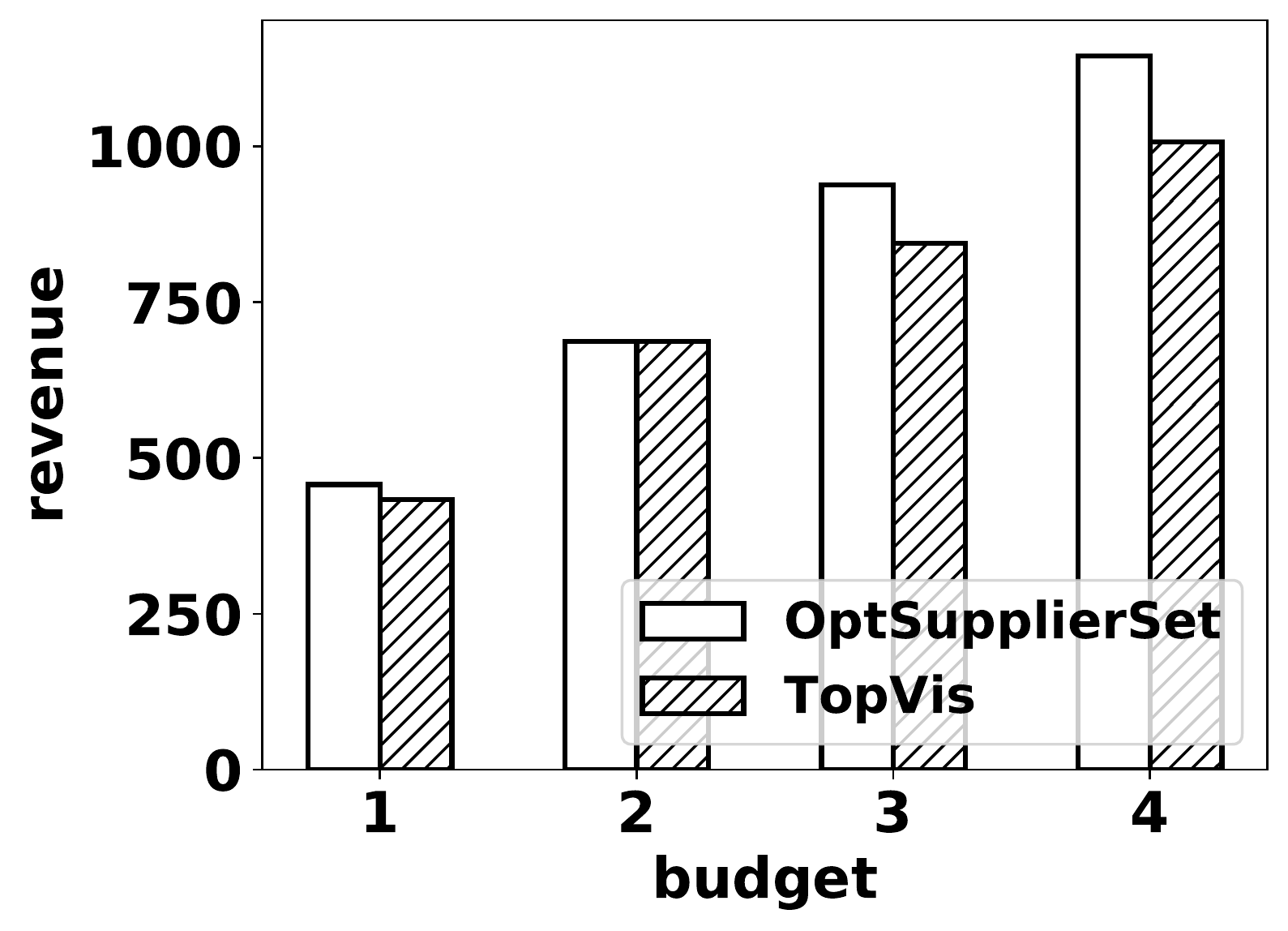}
	}
	\subfigure[DBLP]{
		\includegraphics[width=0.225\textwidth]{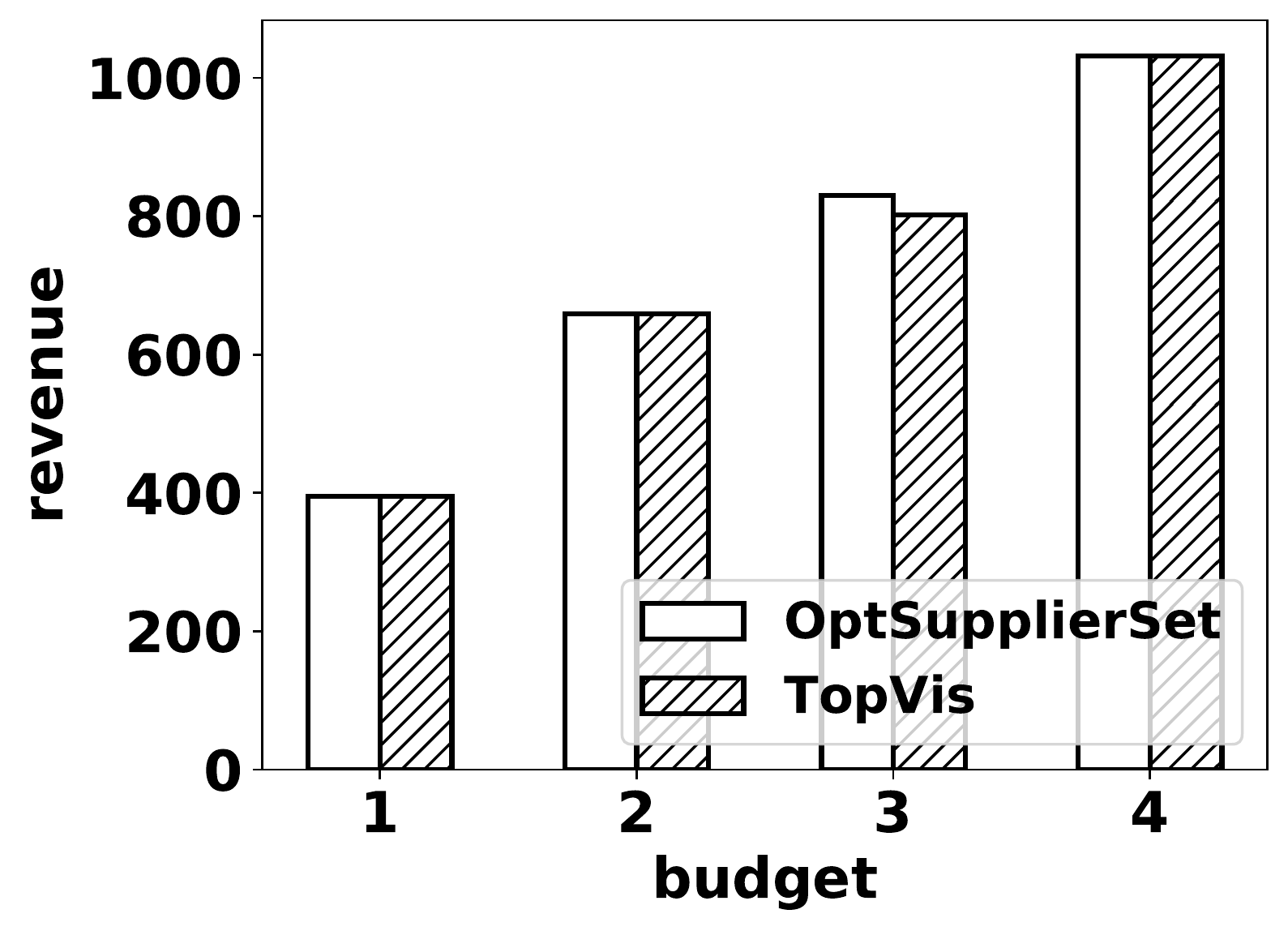}
	}
	\caption{Revenue under different methods to find optimal supplier set,  where discrete search step  $\epsilon = 0.025$.  }
	\label{fig:rev_DicPrice_3Supp}
\end{figure}

We first compare our \texttt{OptSupplierSet} algorithm, i.e., Algorithm \ref{alg:greedy}, 
with two baselines (1) \texttt{Brute} and (2) \texttt{TopVis}.
Note that algorithm \texttt{Brute} is computationally expensive, 
we only experiment it on small datasets Residence and DBLP.
Figure~\ref{fig:rev_DicPrice_3Supp} shows the 
revenue achieved by different methods of finding optimal supplier set, 
where we fix search step as $\epsilon = 0.025$.
From Figure \ref{fig:rev_DicPrice_3Supp}, 
one can observe that the revenue under our 
\texttt{OptSupplierSet} algorithm is nearly the same as that 
under the \texttt{Brute} algorithm on dataset Residence and Blogs.  
This implies a high accuracy of our \texttt{OptSupplierSet} 
in approximating the optimal supplier set.  
One can also observe that the revenue under the heuristic algorithm 
\texttt{TopVis} is slightly smaller than that under our \texttt{OptSupplierSet} on all 4 datasets.  
One reason for this result is that we random select a small set 
of suppliers from the user population, 
which may cause a weak  network externality effect among 
suppliers. 
When this externality effect among suppliers is weak, 
the optimal supplier set becomes roughly the same 
as the set of suppliers with top-$b$  
social visibility.  

\subsection{\bf Evaluating the Discretized Search Algorithm}
Next, we study the impact of search step $\epsilon$ on \texttt{OptPrice}, i.e., Algorithm \ref{alg:DS}.
We vary the search step $\epsilon$ from 
0.2, 0.1, 0.05, 0.025 to 0.0125 on all 4 datasets. 
Besides, we experiment exhaustive search of price on small datasets Residence and Blogs.
Note that if the operator iterates
all values in $\{ p_u | u \in \mathcal{R}  \} \cup \{ q_u /\alpha | u \in \mathcal{S}  \} $ 
as posted price,
then he can iterate all possible combination of participating requesters and potential participating suppliers. 
Thus, exhaustive search of price can be achieved by searching all the values in 
$\{ p_u | u \in \mathcal{R}  \} \cup \{ q_u /\alpha | u \in \mathcal{S}  \} $.
However, it is only practical
when we only have a finite and small set of requesters.  
We did not experiment exhaustive price search on the dataset Facebook and DBLP,
since it is computationally expensive.

\begin{figure}[htb]
	\centering  
	\subfigure[budget $b=1$]{
		\includegraphics[width=0.225\textwidth]{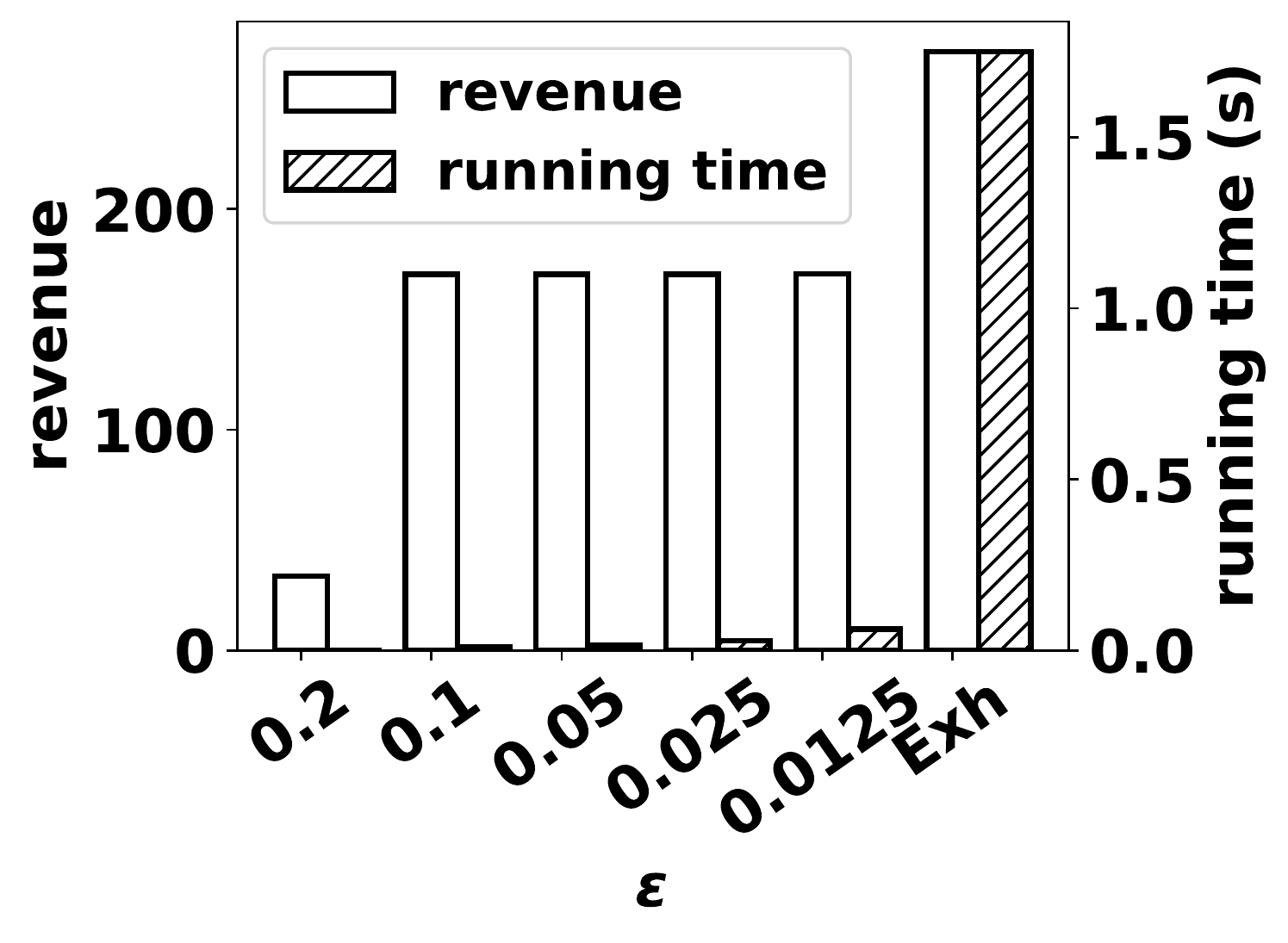}
	}
	\subfigure[budget $b=2$]{
		\includegraphics[width=0.225\textwidth]{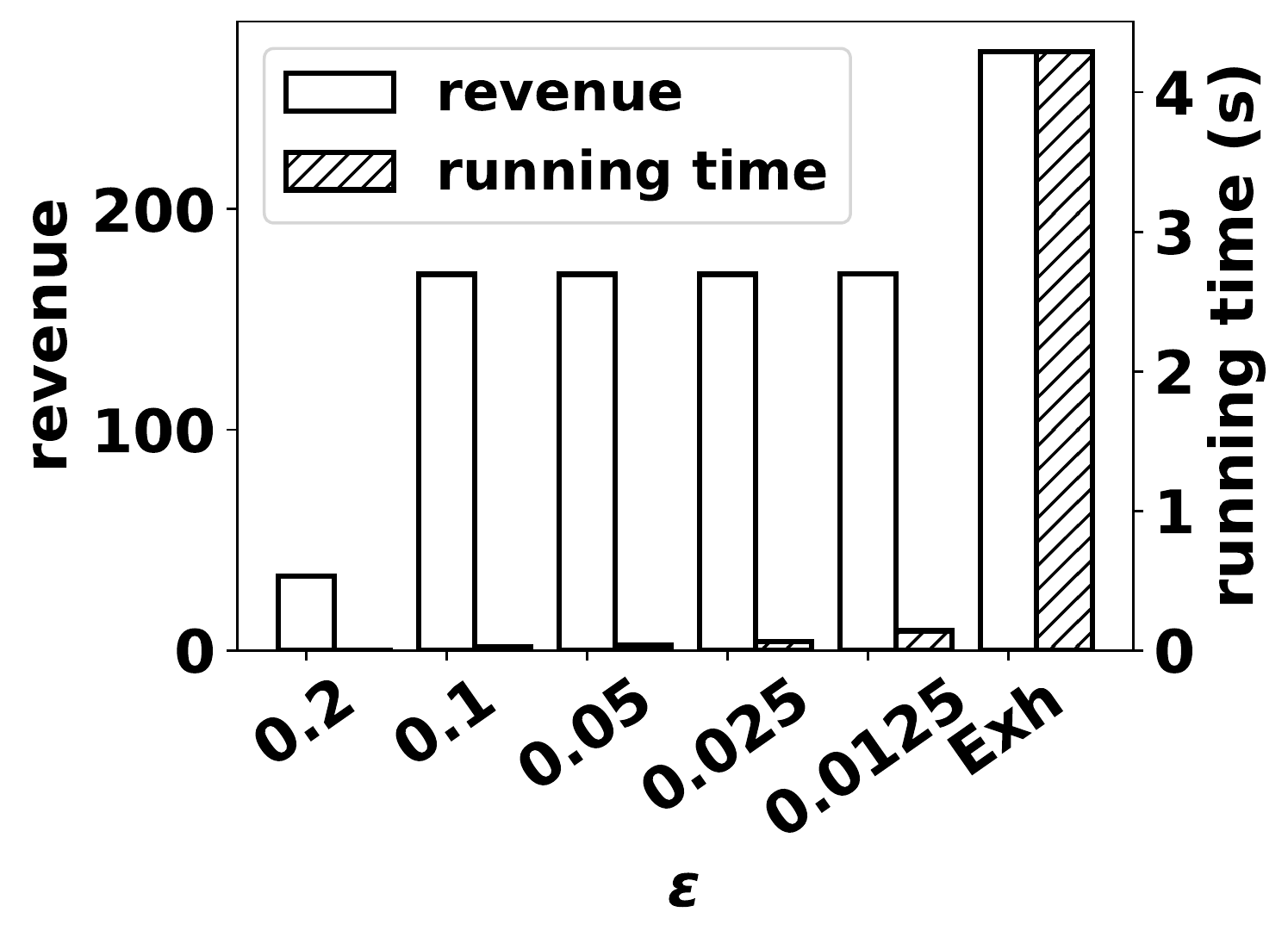}
	}
	\\
	\subfigure[budget $b=3$]{
		\includegraphics[width=0.225\textwidth]{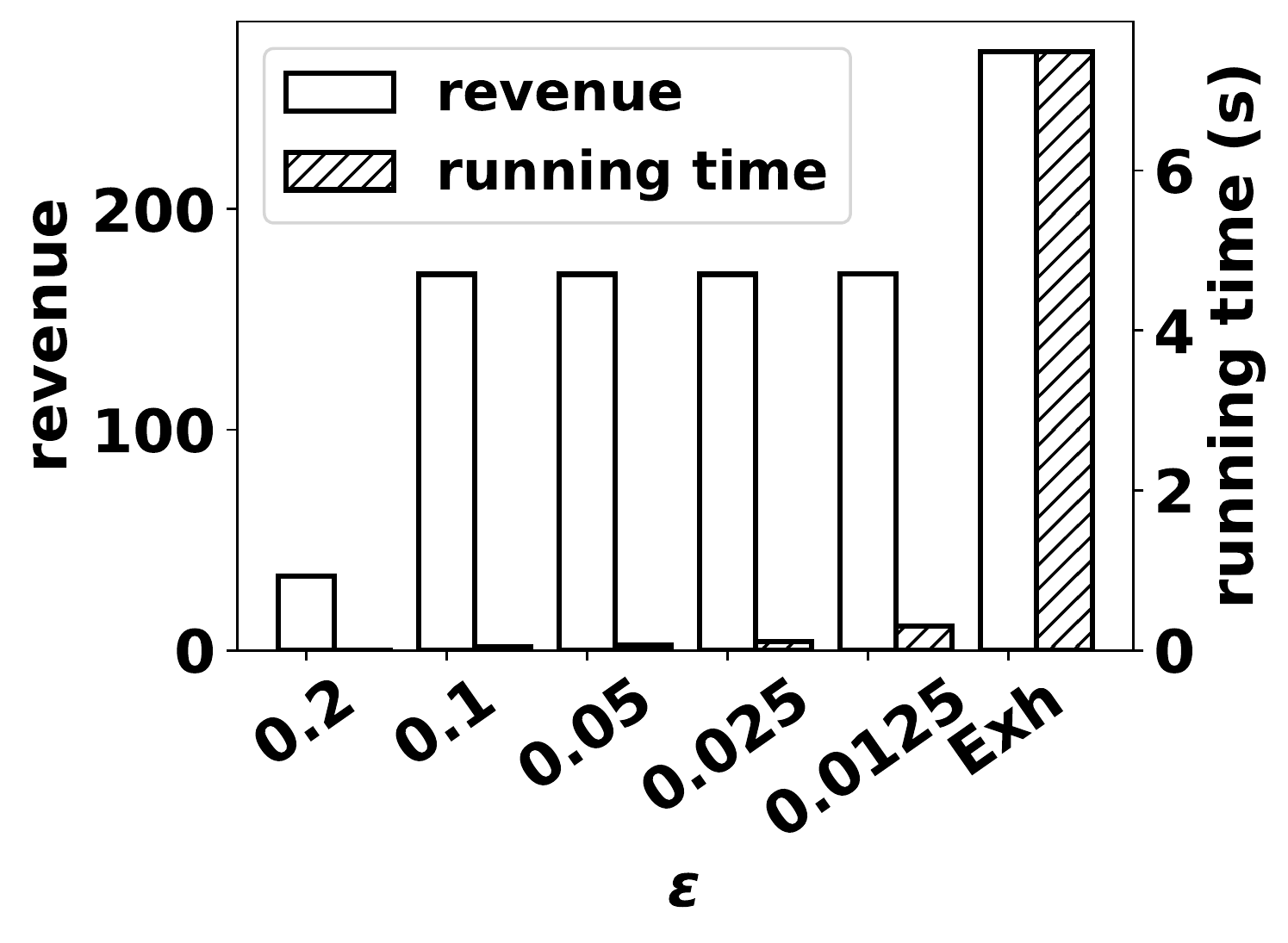}
	}
	\subfigure[budget $b=4$]{
		\includegraphics[width=0.225\textwidth]{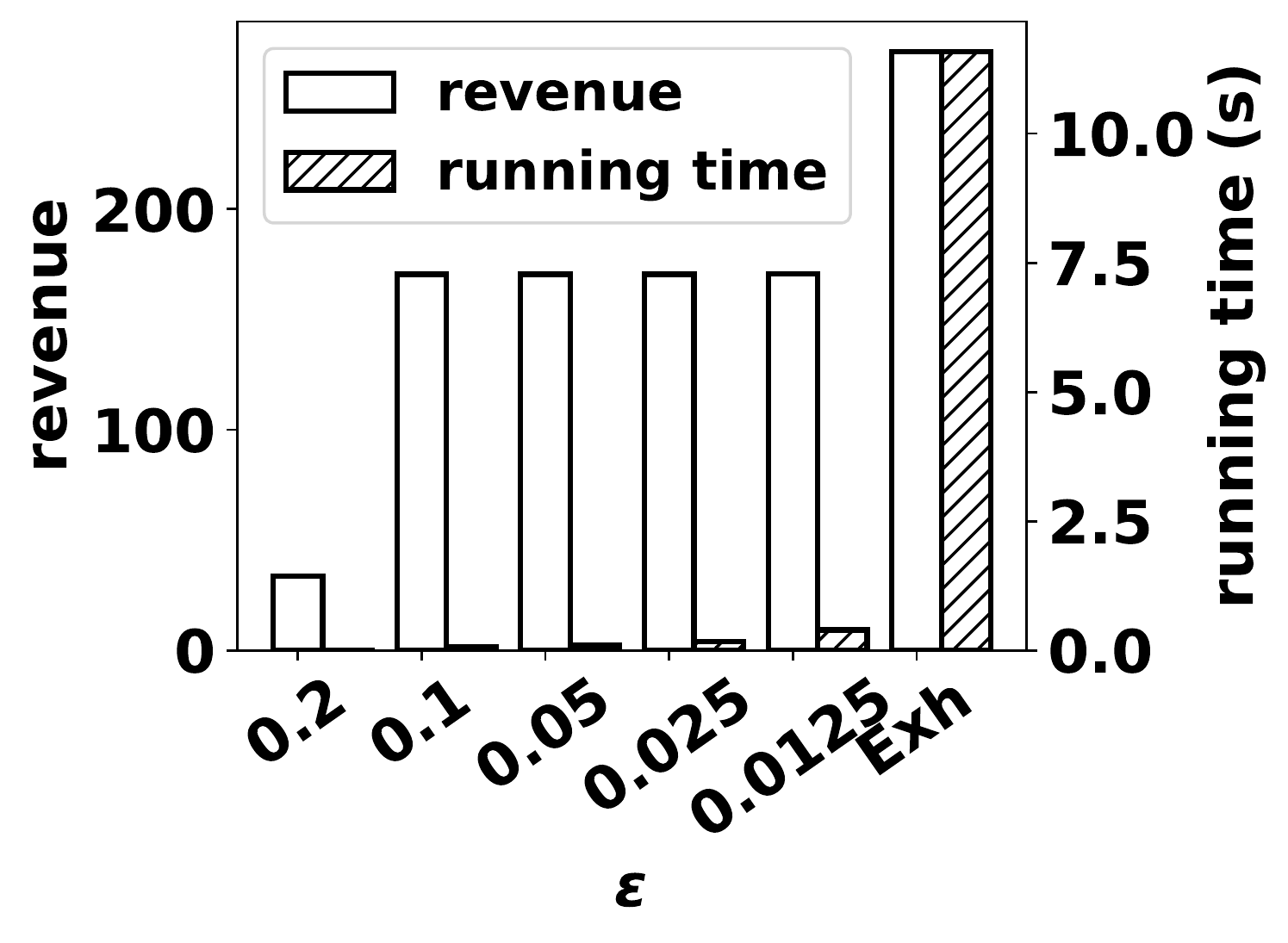}
	}
	\caption{Revenue and running time of \texttt{OptPrice} 
	under different search step size $\epsilon$ (Residence).}
		\label{fig:grain_Price_AdaSupp_residence}
\end{figure}

\begin{figure}[htb]
	\centering  
	\subfigure[budget $b=1$]{
		\includegraphics[width=0.225\textwidth]{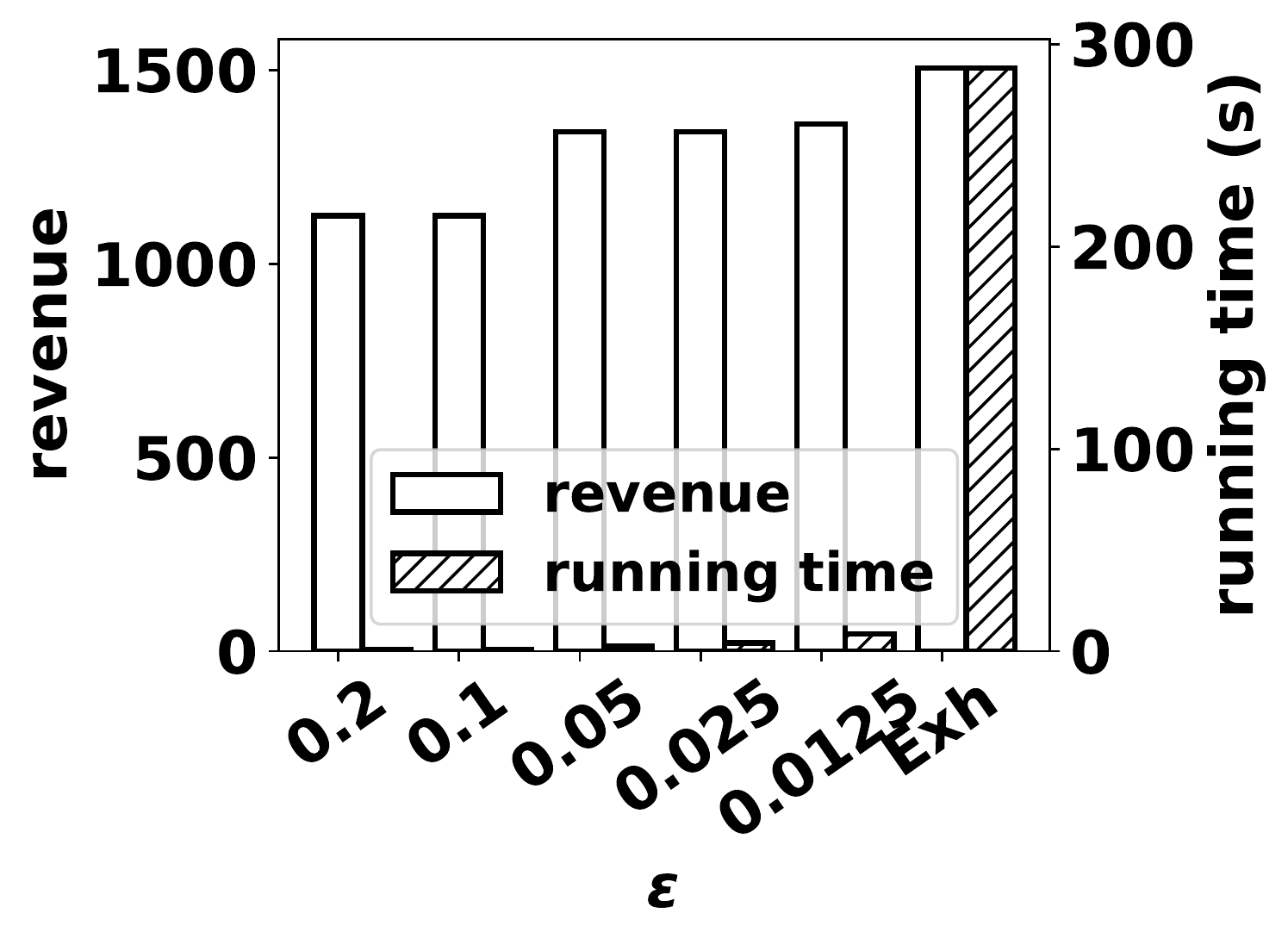}
	}
	\subfigure[budget $b=2$]{
		\includegraphics[width=0.225\textwidth]{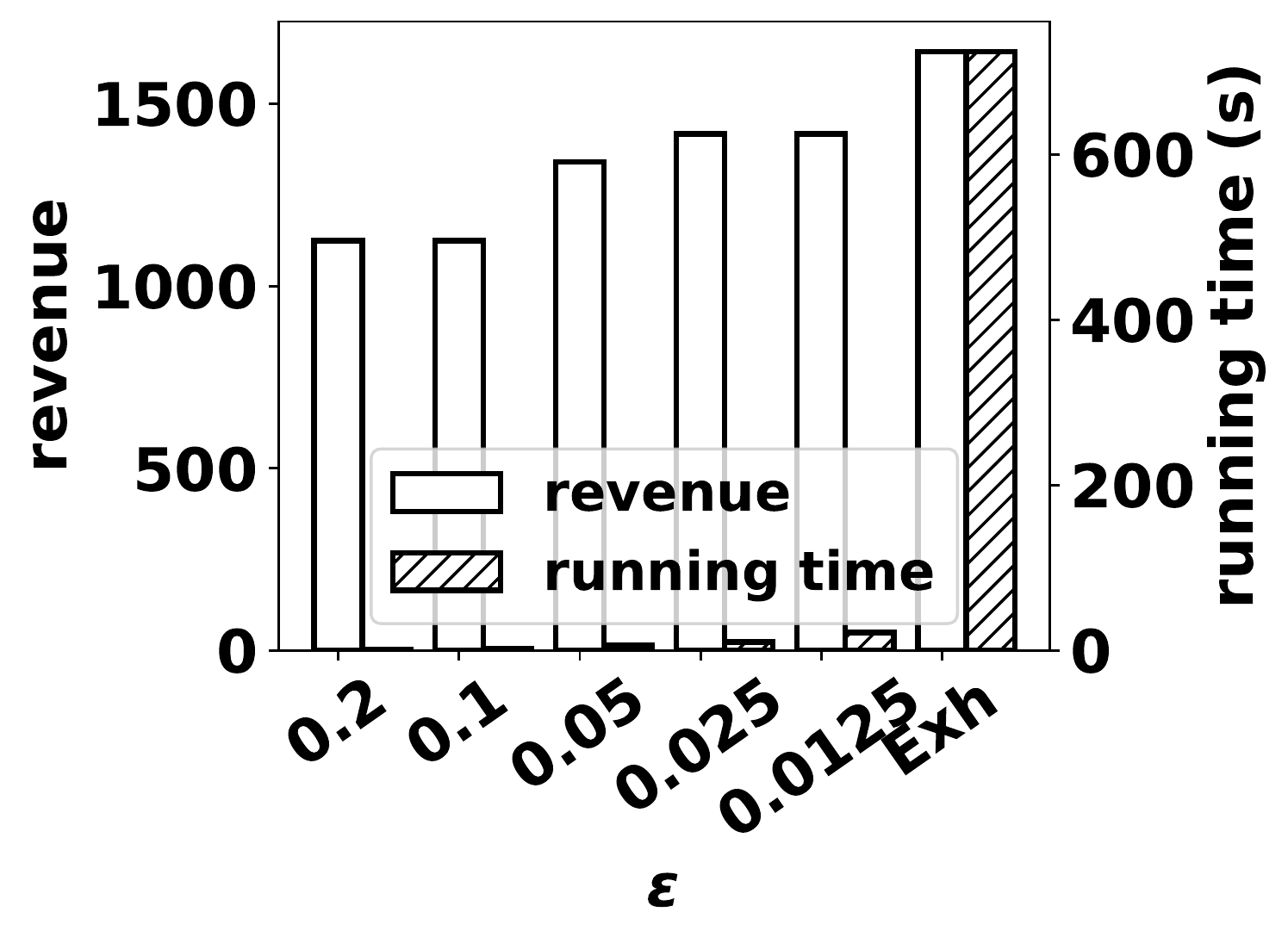}
	}
	\\
	\subfigure[budget $b=3$]{
		\includegraphics[width=0.225\textwidth]{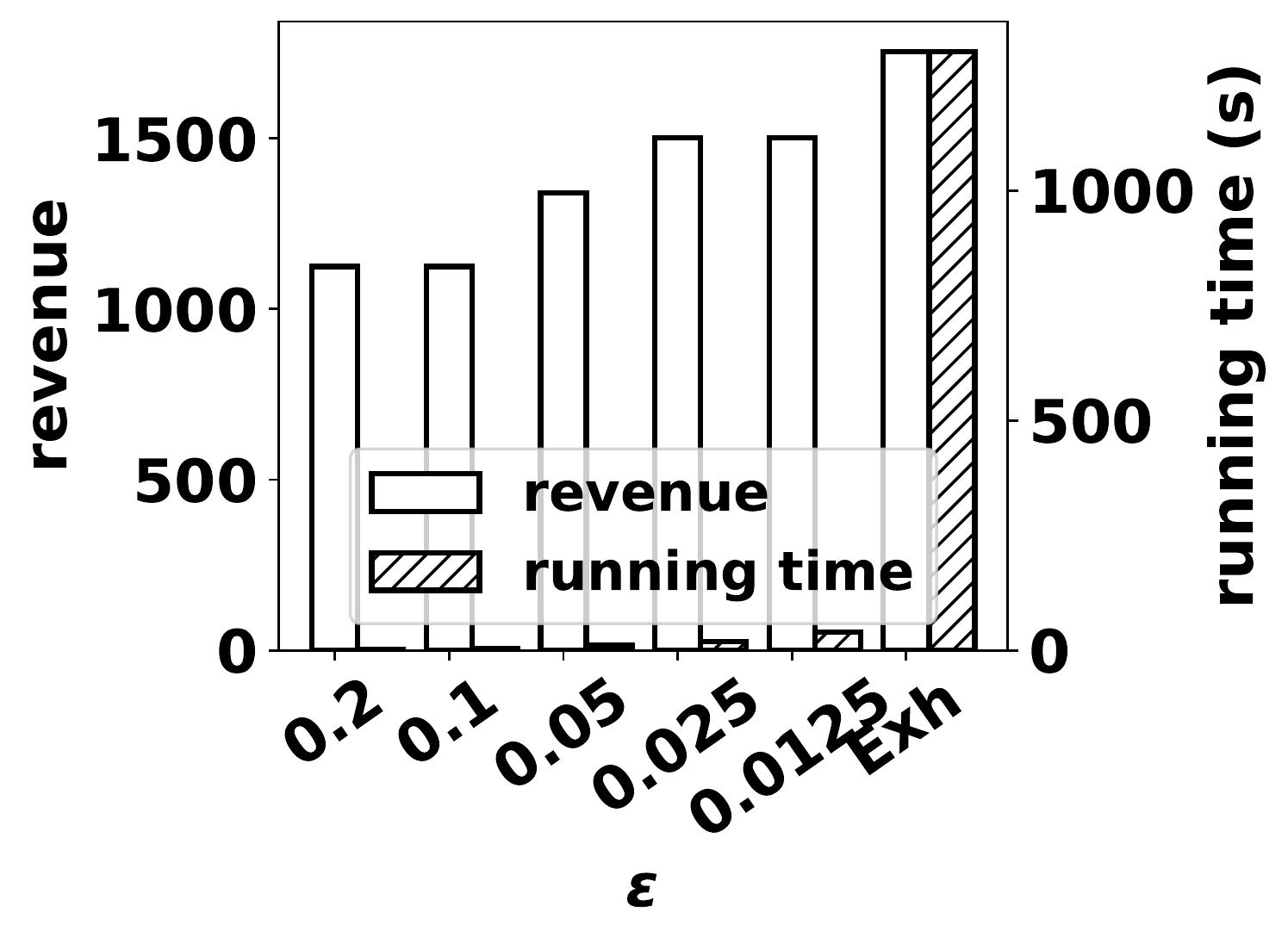}
	}
	\subfigure[budget $b=4$]{
		\includegraphics[width=0.225\textwidth]{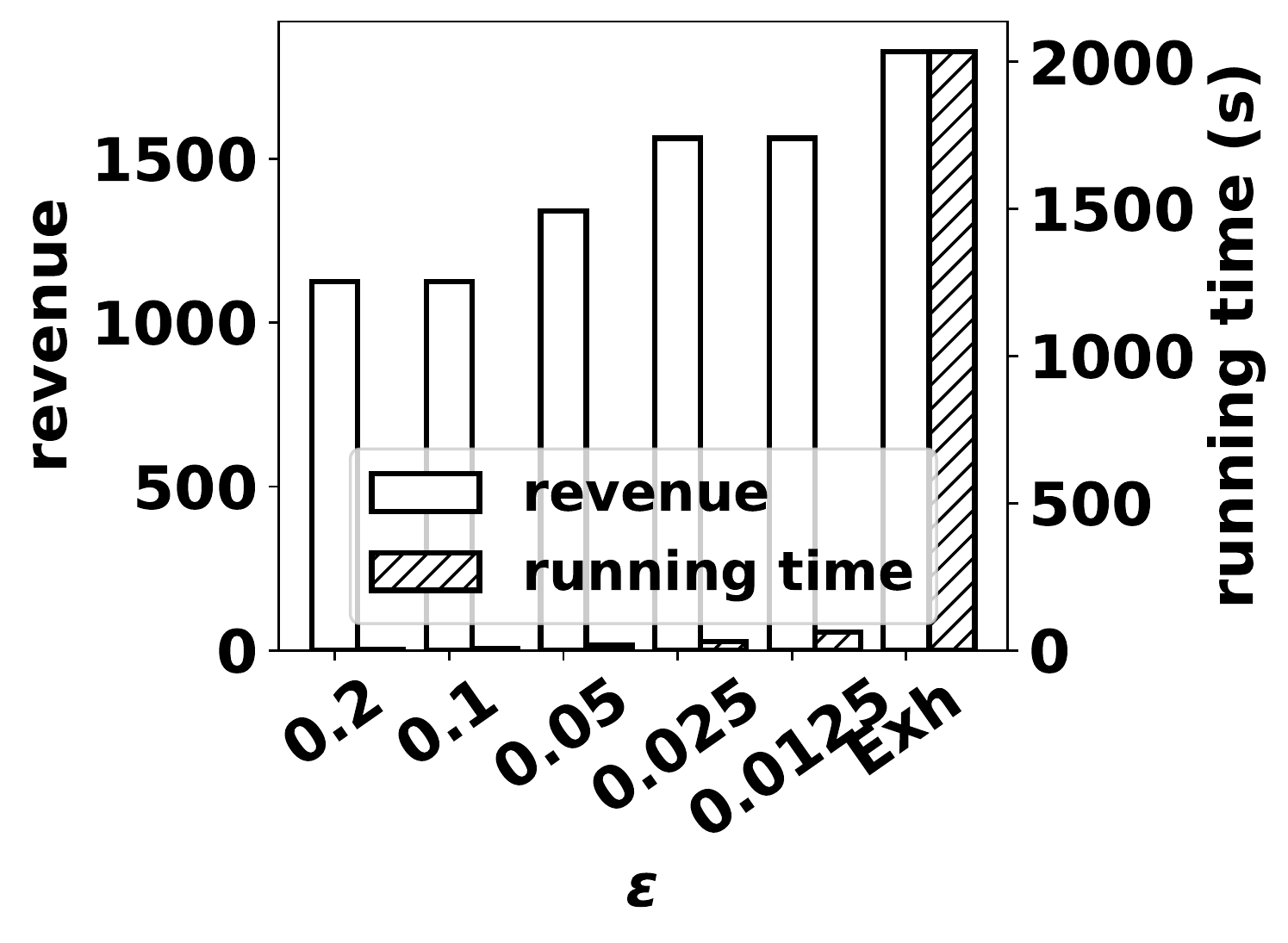}
	}
	\caption{Revenue and running time of \texttt{OptPrice} 
	under different search step size $\epsilon$ (Blogs).}
	\label{fig:grain_Price_AdaSupp_blog}
\end{figure}

\begin{figure}[htb]
	\centering  
	\subfigure[budget $b=1$]{
		\includegraphics[width=0.225\textwidth]{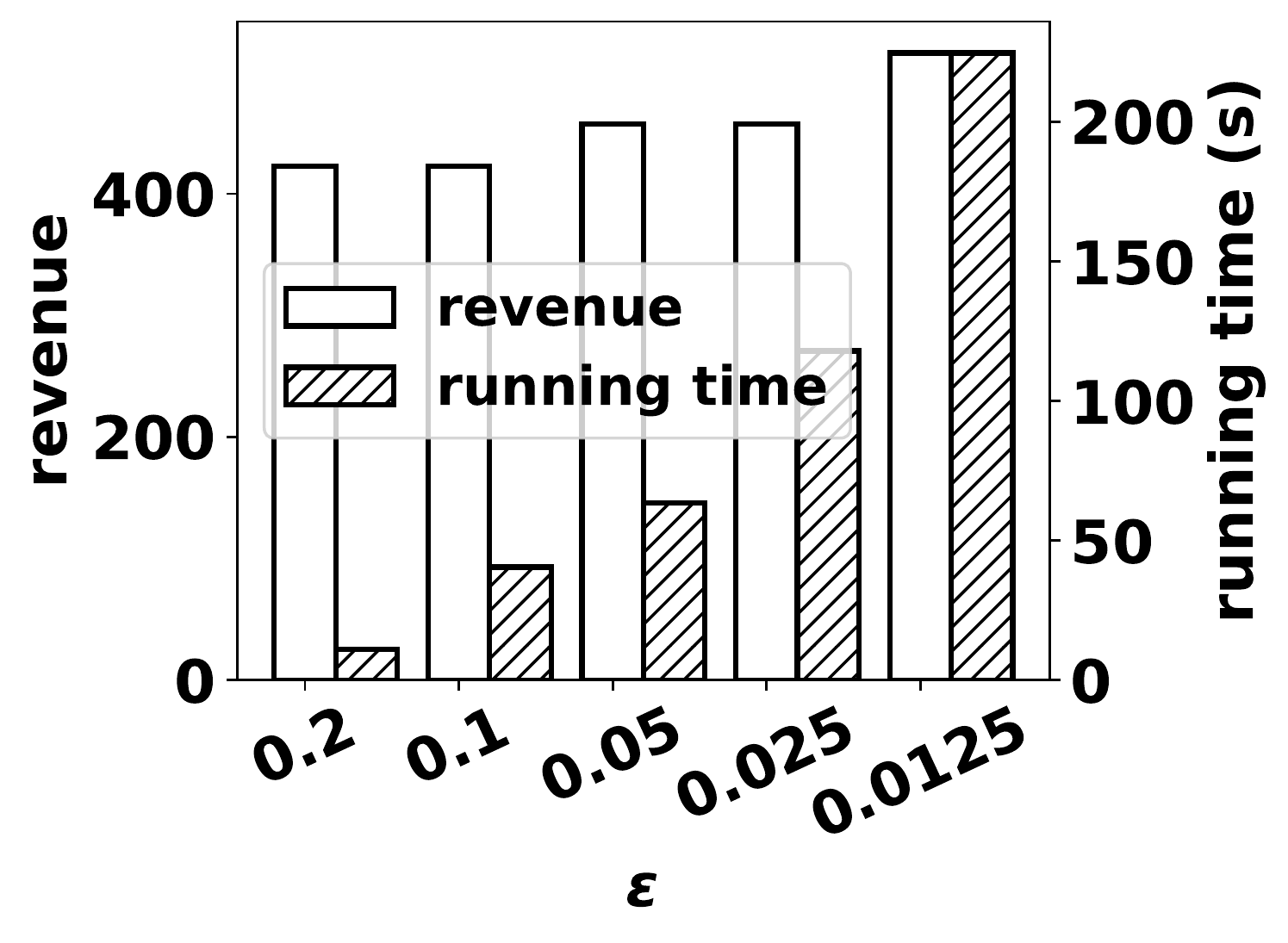}
	}
	\subfigure[budget $b=2$]{
		\includegraphics[width=0.225\textwidth]{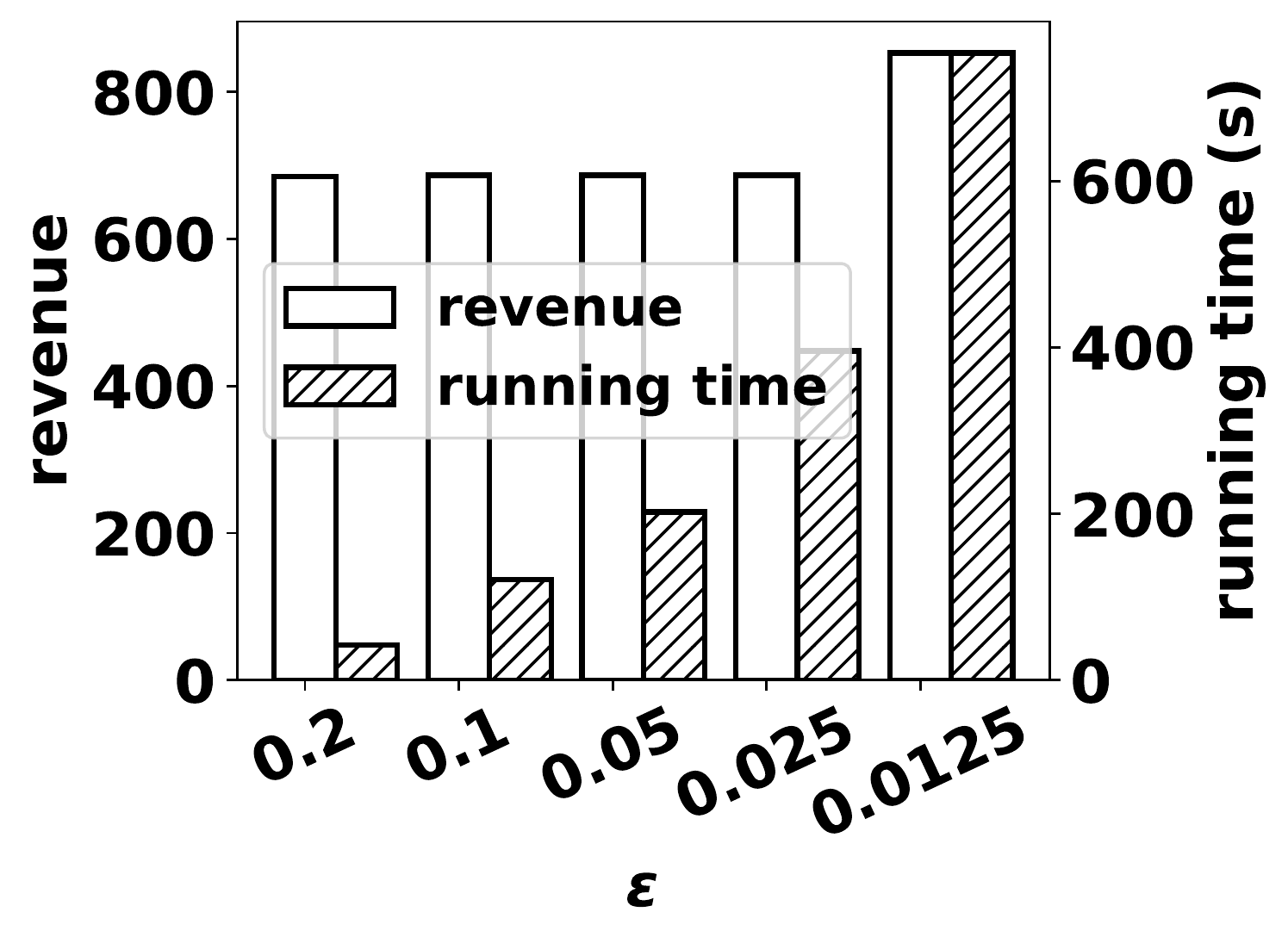}
	}
	\\
	\subfigure[budget $b=3$]{
		\includegraphics[width=0.225\textwidth]{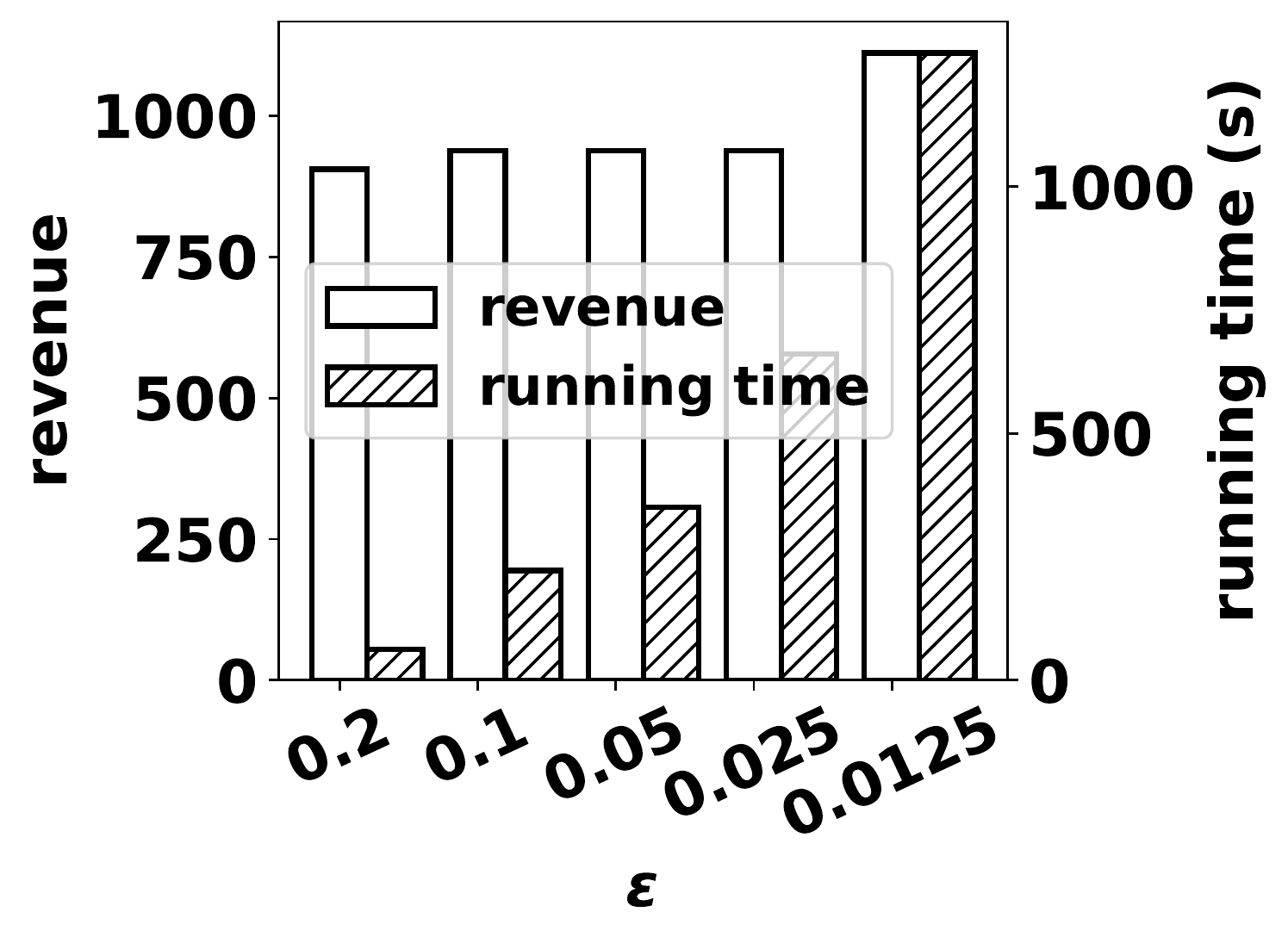}
	}
	\subfigure[budget $b=4$]{
		\includegraphics[width=0.225\textwidth]{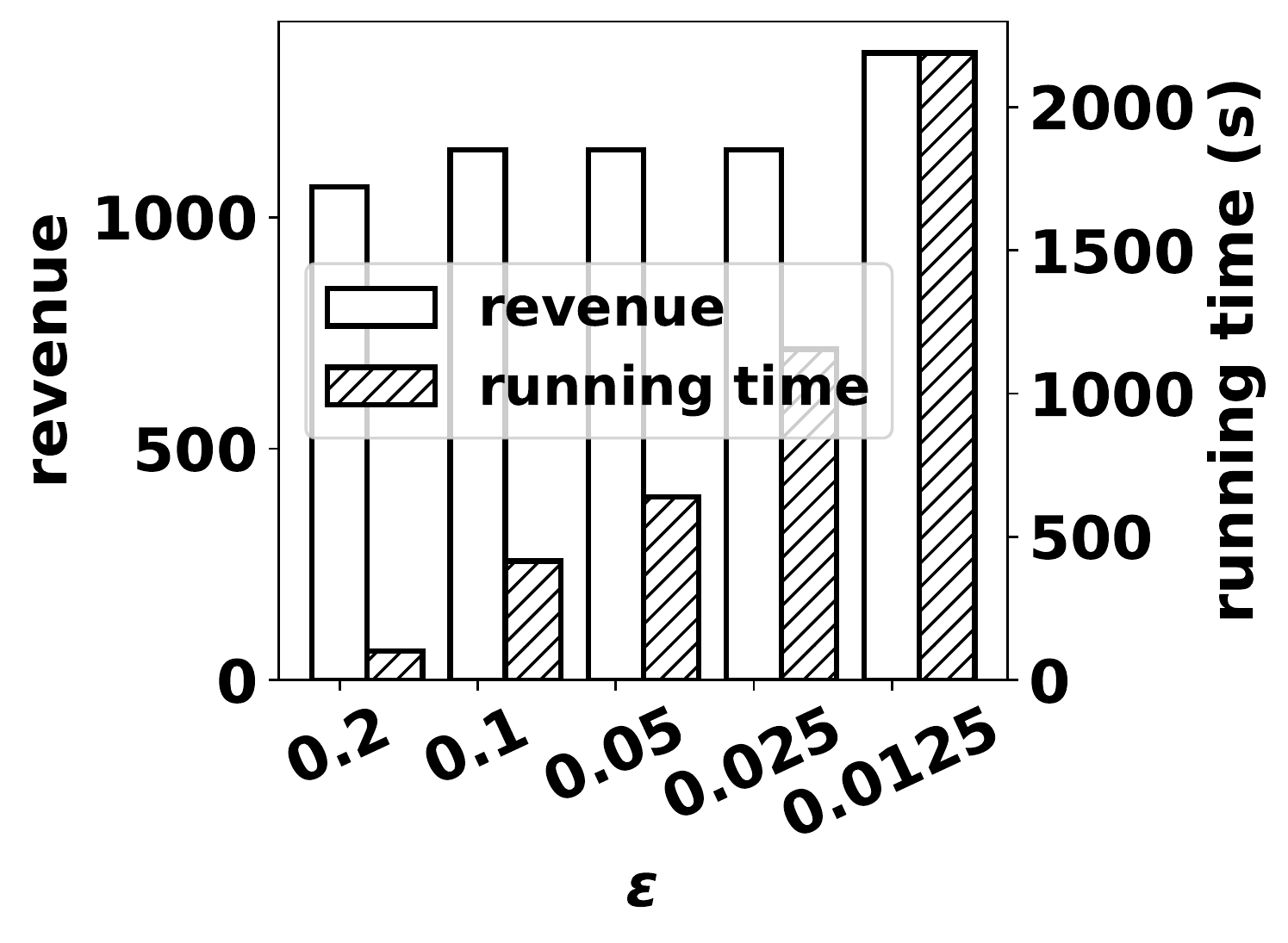}
	}
	\caption{Revenue and running time of \texttt{OptPrice} 
		under different search step size $\epsilon$ (Facebook).}
		\label{fig:grain_Price_AdaSupp_facebook}
\end{figure}

\begin{figure}[htb]
	\centering  
	\subfigure[budget $b=1$]{
		\includegraphics[width=0.225\textwidth]{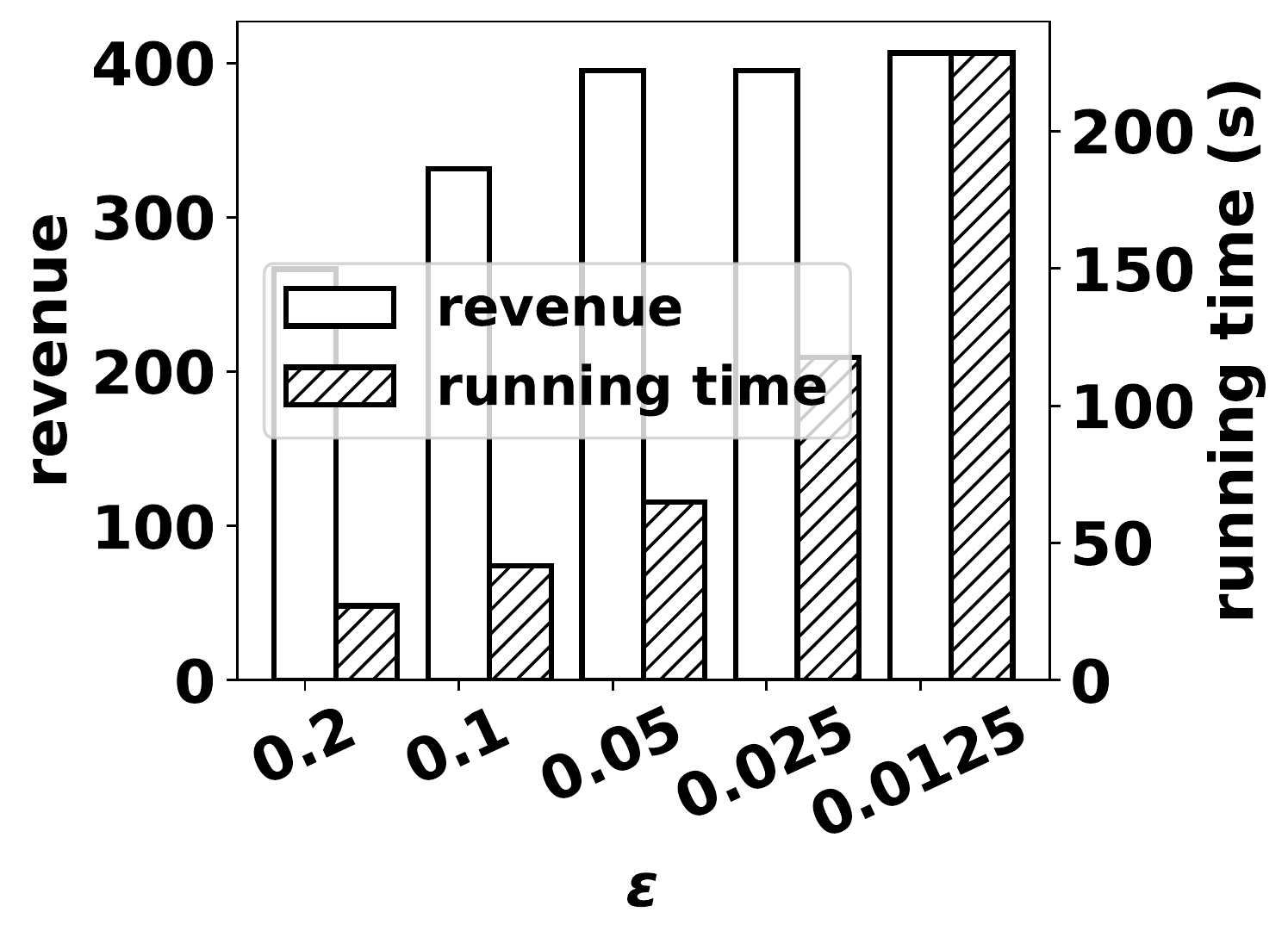}
	}
	\subfigure[budget $b=2$]{
		\includegraphics[width=0.225\textwidth]{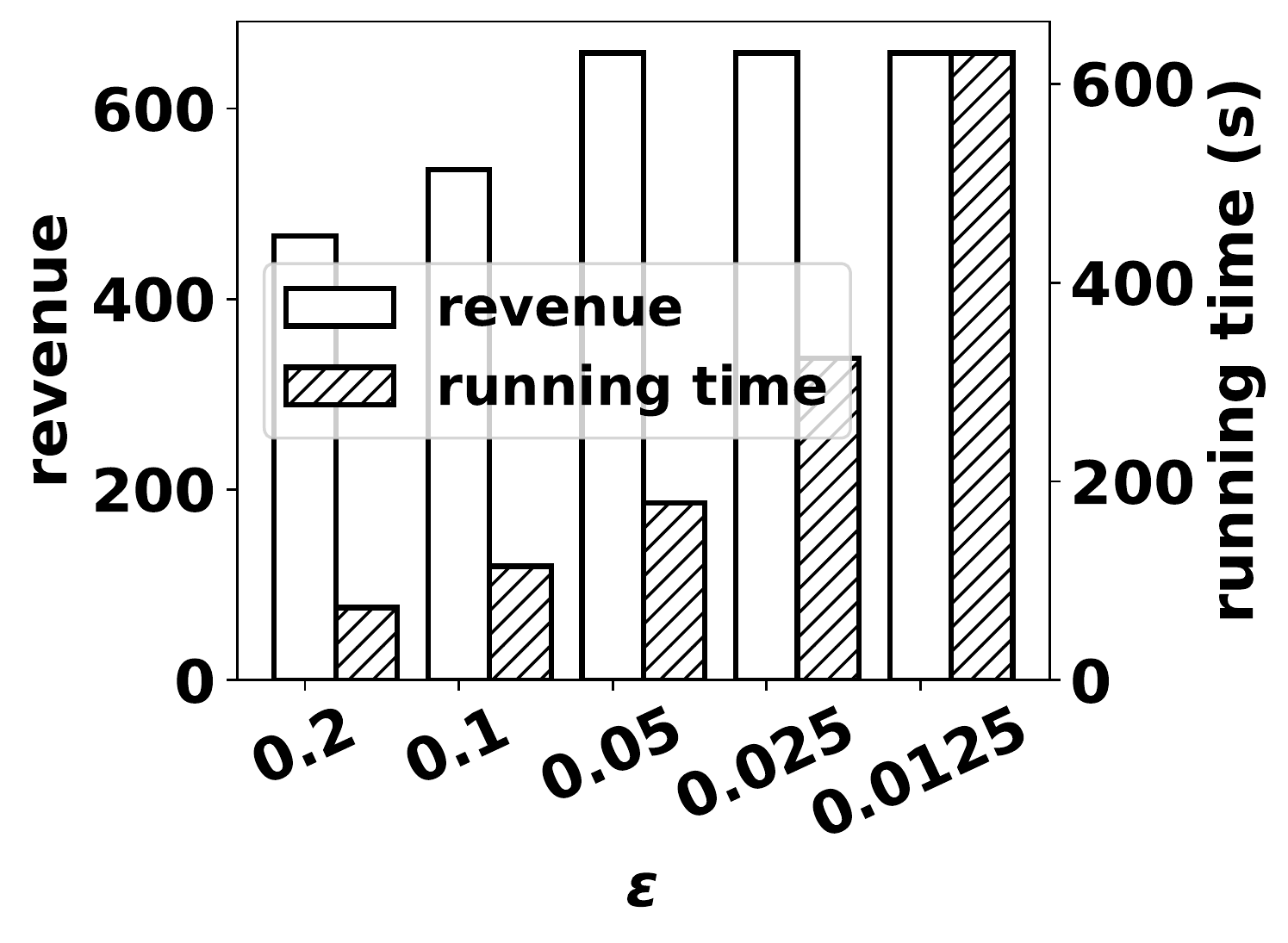}
	}
	\\
	\subfigure[budget $b=3$]{
		\includegraphics[width=0.225\textwidth]{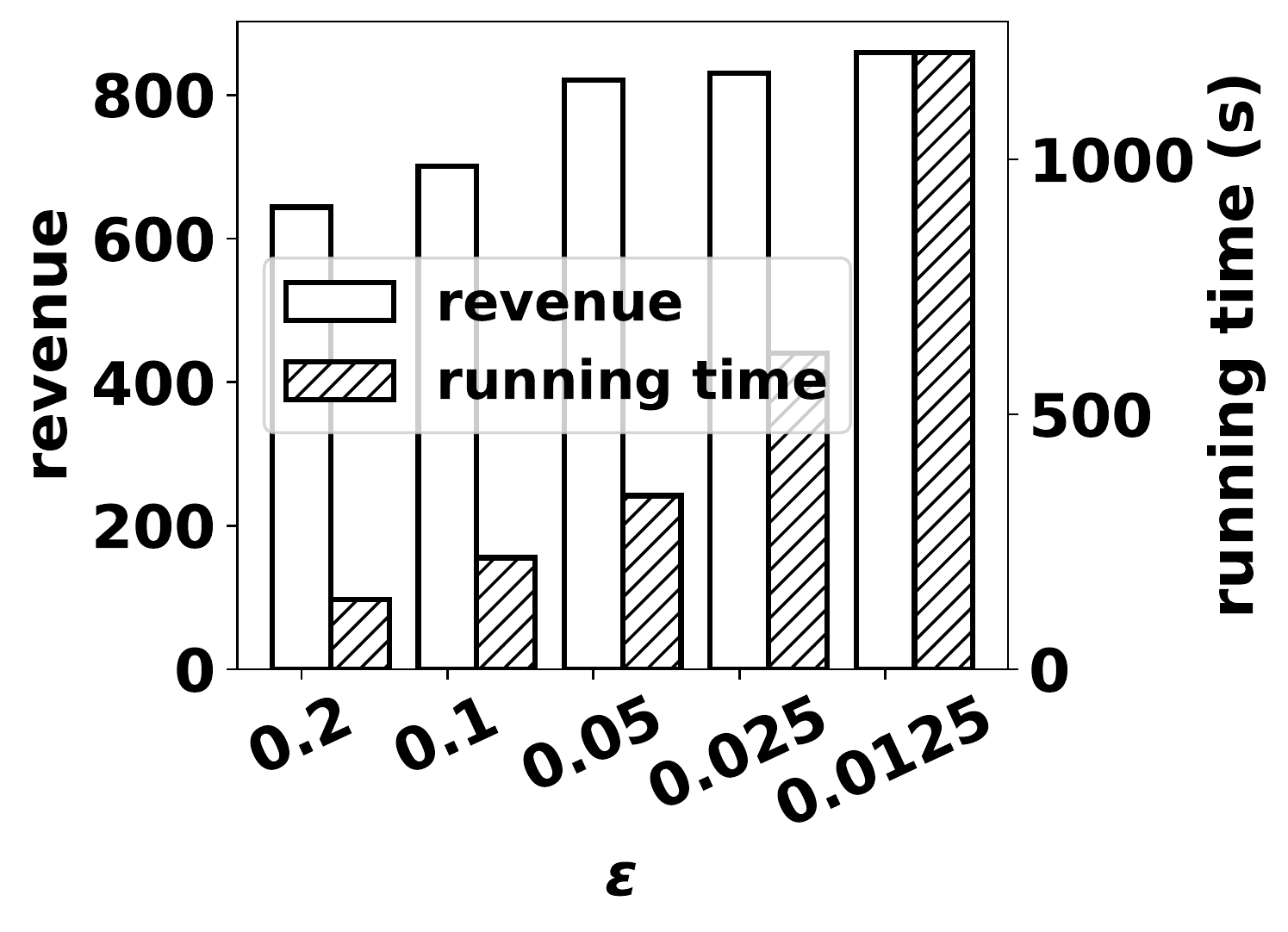}
	}
	\subfigure[budget $b=4$]{
		\includegraphics[width=0.225\textwidth]{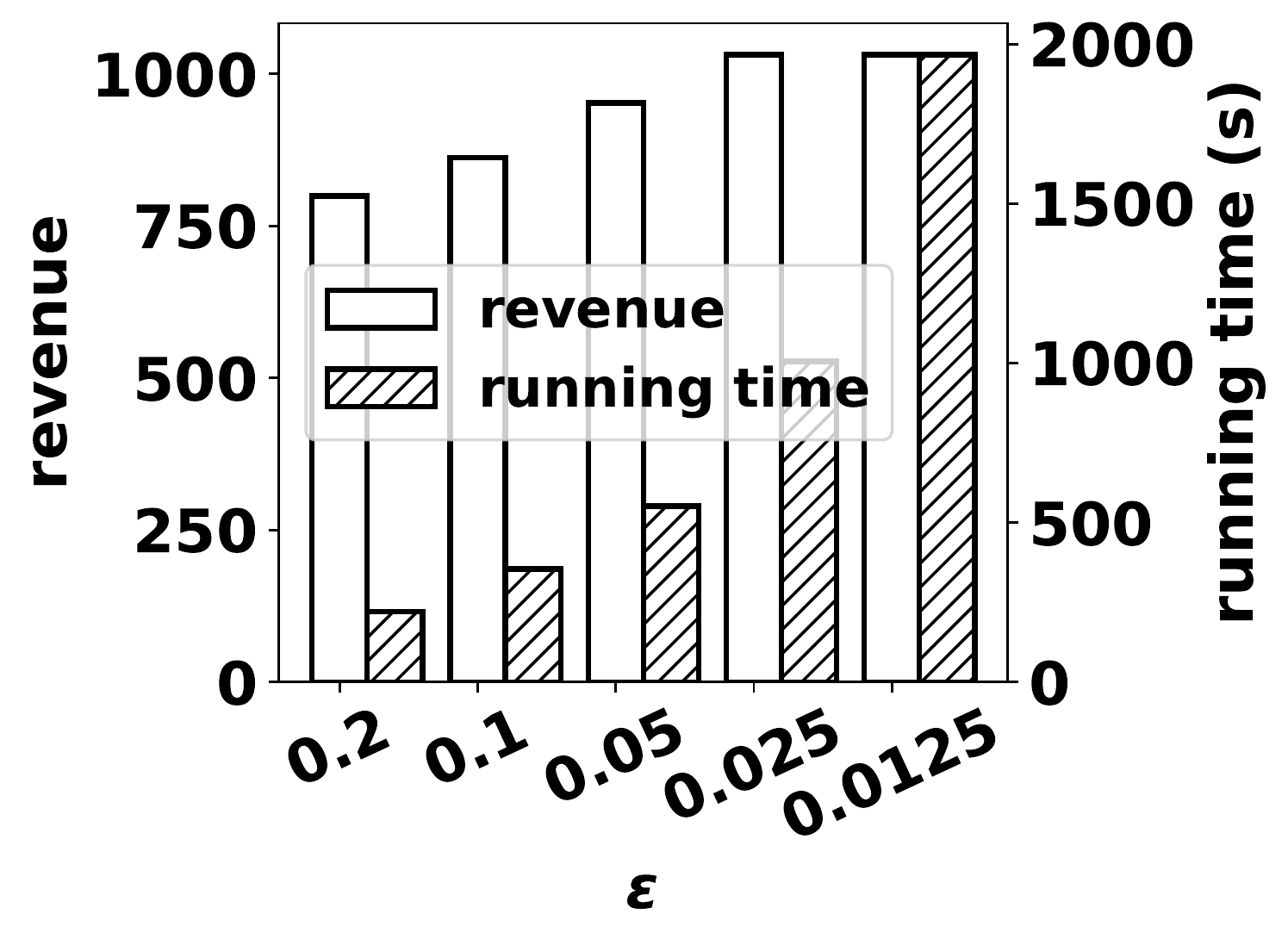}
	}
	\caption{Revenue and running time of \texttt{OptPrice} 
		under different search step size $\epsilon$ (DBLP).}
	\label{fig:grain_Price_AdaSupp_dblp}
\end{figure}

Figure~\ref{fig:grain_Price_AdaSupp_residence}
shows the revenue and running time of \texttt{OptPrice} 
under different selection of search step $\epsilon$ (including exhaustive search of price)
and different budgets $b=1,2,3,4$ on dataset Residence.  
From Figure~\ref{fig:grain_Price_AdaSupp_residence} one can observe that 
on the dataset Residence, 
decreasing the search step $\epsilon$ can increase the revenue of 
\texttt{OptPrice}, but it also increases the running time.  
Moreover, there is a diminishing increase of revenue.   
We can find that $\epsilon = 0.1$ is a 
proper price search step for the dataset Residence,
since it achieves a good balance between running time and performance.  
The bar labeled as \texttt{Exh} corresponds to exhaustive search of price.  
On can observe that with a search step $\epsilon = 0.1$, 
\texttt{OptPrice} can achieve a revenue around 70\% of that under exhaustive search of price, 
which takes at least more than ten times of the running time 
of \texttt{OptPrice}.  
Figure~\ref{fig:grain_Price_AdaSupp_blog} shows the results on dataset Blogs.
The result is similar to that of the dataset Residence. 
However, for the dataset Blogs, we can find that $\epsilon = 0.025$ 
is a proper price search step.
Figure~\ref{fig:grain_Price_AdaSupp_facebook} show the results on dataset Facebook.
We can find  
when we vary search step, 
the revenues achieved do not change much.
Figure~\ref{fig:grain_Price_AdaSupp_dblp} shows the results on dataset DBLP.
One can observe that 0.05 is a proper search step.


%% file: 6_related.tex

\section{\bf Related Work}
\label{sec:related_work} 

The notion of social visibility defined in this paper 
is closely related to social influence   \cite{kempe2003maximizing,domingos2001mining,Lin2017,richardson2002mining}.   
One key difference is that when a user is visible to a set of users, 
it does not mean this user can influence this set of users.  
The objective of influence maximization problem 
is to find a subset of nodes 
that could maximize the spread of information 
under certain influence diffusion models.  
But our problem focuses on the pricing of social visibility service.  
The idea of adding new links to enhance social visibility 
is closely related 
to link prediction 
\cite{L__2011, martinez2017survey,liben2007link}, 
and friend recommendation  \cite{resnick1997recommender,ma2014measuring,ma2011recommender,xie2010potential}, The objectives of link prediction and friend recommendation 
are to predict future or missing links.  
Our work adds links that can improve social visibility.  
Note that such links may have nothing to do with predicting the future 
or missing links.  

Technically, our work is closely related to revenue management \cite{Talluri2004}.  
Different with classical revenue management literatures \cite{Talluri2004}, 
which focuses on understanding the structure of optimal pricing, 
our work formulates a new revenue maximization framework and we focus 
design approximation algorithms to solve this problem.  
Similar to influence maximization  \cite{kempe2003maximizing,domingos2001mining,richardson2002mining}, 
the core technique in selecting the supplier set is submodular analysis.  
Our contribution is in proving that our problem has the submodular property 
and show that how the submodular property impacts the search 
optimal pricing.

%% file: 7_conclusions.tex

\section{\bf Conclusions}
\label{sec:conlcusions}

This paper proposes a posted pricing scheme for 
the OSN operator to price its social visibility boosting service.    
We formulate a revenue maximization problem for 
the OSN provider to select the parameter of the posted pricing scheme.  
We show that the revenue maximization problem is not simpler 
than an NP-hard problem.  
We decomposed it into two sub-routines, 
where one focuses on selecting the optimal set of suppliers, 
and the other one focuses on selecting the optimal prices.  
We prove the hardness of each sub-routine, 
and eventually design a computationally efficient approximation algorithm to solve 
the revenue maximization problem with provable theoretical guarantee 
on the revenue gap.  
We conduct extensive experiments on four public datasets to validate 
the superior performance of our proposed algorithms.

%
%
%
%
%
%